\documentclass[12pt]{article}

\usepackage{amsmath}
\usepackage{graphicx,psfrag,epsf}
\usepackage{enumerate}
\usepackage{natbib}
\usepackage{url} 
\usepackage{booktabs}
\usepackage{verbatim}

\newcommand{\var}{\mbox{Var}}
\newcommand{\corr}{\mbox{Corr}}
\newcommand{\bbeta}{\mbox{\boldmath $\beta$}}
\newcommand{\bmu}{\mbox{\boldmath $\mu$}}
\newcommand{\bpi}{\mbox{\boldmath $\pi$}}
\newcommand{\bSigma}{\mbox{\boldmath $\Sigma$}}
\newcommand{\btheta}{\mbox{\boldmath $\theta$}}

\newcommand{\boeta}{\mbox{\boldmath $\eta$}}
\newcommand{\bTheta}{\mbox{\boldmath $\Theta$}}
\newcommand{\colvec}[2][.9]{%
  \scalebox{#1}{%
    \renewcommand{\arraystretch}{.8}%
    $\begin{bmatrix}#2\end{bmatrix}$%
  }
}
\newcommand{\blind}{0}

\begin{document}

\def\spacingset#1{\renewcommand{\baselinestretch}%
{#1}\small\normalsize} \spacingset{1}


\if0\blind
{
\title{\bf AR(1) Latent Class Models for Longitudinal Count Data}
\author{Nicholas C.\ Henderson\thanks{
    Nicholas C.\ Henderson is PhD Candidate, Department of Statistics,
    University of Wisconsin, Madison, WI 53792 (e-mail: nhenders@stat.wisc.edu).
    Paul J.\ Rathouz is Professor, Department of Biostatistics and Medical
    Informatics, University of Wisconsin, Madison, WI 53792 (e-mail: rathouz@biostat.wisc.edu). 
    Henderson's contributions to this work were supported
    by 5T32HL083806 from the National Heart, Lung, and Blood Institute. Rathouz'
    contributions were supported by grant 5R01HD061384 from the National
    Institute of Child Health and Human Development.} \hspace{.2cm} \\
Department of Statistics \\
University of Wisconsin-Madison 
\and
Paul J.\ Rathouz \\
Department of Biostatistics \& Medical Informatics \\
University of Wisconsin-Madison }
\date{}
\maketitle
} \fi

\if1\blind
{
  \bigskip
  \bigskip
  \bigskip
  \begin{center}
    {\LARGE\bf  AR(1) Latent Class Models for Longitudinal Count Data}
\end{center}
  \medskip
} \fi

\bigskip
\begin{abstract}
  In a variety of applications involving longitudinal or
  repeated-measurements data, it is desired to uncover natural groupings
  or clusters which exist among study subjects. Motivated by the need to
  recover longitudinal trajectories of conduct problems in the field of
  developmental psychopathology, we propose a method to address this goal
  when the data in question are counts. We assume that the
  subject-specific observations are generated from a first-order
  autoregressive process which is appropriate for counts. A key advantage
  of our approach is that the marginal distribution of the response can be
  expressed in closed form, circumventing common computational issues
  associated with random effects models. To further improve computational
  efficiency, we propose a quasi-EM procedure for estimating the model
  parameters where, within each EM iteration, the maximization step is
  approximated by solving an appropriately chosen set of estimating
  equations. Additionally, we introduce a novel method to express the
  degree to which both the underlying data and the fitted model are able
  to correctly assign subjects to their (unknown) latent classes. We
  explore the effectiveness of our procedures through simulations based on
  a four-class model, placing a special emphasis on posterior
  classification.  Finally, we analyze data and recover trajectories of
  conduct problems in an important nationally representative sample.
\end{abstract}

\noindent%
  \vfill

\newpage
\spacingset{1.45} 
\section{Introduction}
\label{sec:intro}

Many longitudinal studies have the aim of tracking the change in some
outcome or response over time.  This is an important and common goal in
the field of developmental psychopathology, which aims to study the
natural history of common childhood psychiatric diseases such as conduct
disorder and delinquency.  Often, there exists substantial variability in
the response-paths of observed trajectories across subjects, and grouping
subjects with similar trajectories may reveal certain sub-populations that
exhibit interesting developmental patterns.  In conduct disorder research,
for example, 
such distinct ``developmental sub-types'' of trajectories are of great
interest because the classification carries important information about level of impairment, future life
outcomes, and possible etiologic origin (see, e.g. \citet{Odgers:2007}, or
\cite{Moffitt:1993}). Furthermore, it is of interest to robustly estimate
such trajectory sub-types using large and representative samples, to do so
in a computationally efficient way, and to use those estimates to recover
class membership at the subject level.   The problem of identifying a
finite number of sub-populations is frequently formulated as a latent
class or finite mixture model where the distribution governing the
observations on a given subject is determined by an unobserved class
label.

In an alternative approach to the analysis of longitudinal data, random effects are introduced to account for the heterogeneity
across subjects and the correlation among observations on the same subject. 
If the conditional distribution of the response given the values of the random effects is not
Gaussian however, the marginal distribution of the response will typically not have a closed form.
In these cases, evaluation of the likelihood requires numerical integration over the distribution
of the random effects. Direct maximization of the likelihood then involves numerical integration
for every evaluation of the likelihood.
A number of other estimation approaches for models of this type, commonly referred to
as generalized linear mixed models (GLMMs), have been proposed, including
approximate methods such as penalized quasi-likelihood (\cite{Schall:1991} or \cite{Breslow:1993}),
Monte Carlo methods (\cite{Zeger:1991}), and marginalized random effects models
(\cite{Heagerty:1999}).

More recently, some authors have combined these two approaches. They have
observed that with longitudinal data, a small number of latent classes is
not sufficient to account for all of the realized between-subject
heterogeneity and correlation among observations within subjects. They
have therefore developed models with latent random effects in addition to
a latent variable indicating class membership already present (see
e.g. \cite{Muthen:1999}, or \citet{Qu:1996}). Whereas this approach has
gained traction in the applied literature, it poses two potential
methodological problems. First, the addition of within-class random
effects to the latent class model may complicate computation considerably.
For example, if using an EM algorithm for estimation, not only does one
confront within each iteration the difficulties associated with GLMMs, but
one must also use numerical integration to update the class-membership
probabilities within each iteration. A within-class \emph{closed form}
likelihood would be much more computationally tractable.

The second problem involves the distinction between global and local
correlation among observations on the same subject.  To articulate this
concern, we first posit that one key role of the latent class structure is
to account for the \emph{global} correlation, i.e., the correlation that
exists between all observations on a subject, whether they are separated
by a short or by a long time lag. In classic latent class analysis, given
class membership, all observations on a given subject are independent, and
this assumption drives the identification of the model. This assumption
is, however, somewhat restrictive and there are concerns that it could
lead to models with too many classes that are too small if there remains
residual correlation among some of the observations within subject. An
obvious solution is to allow---and model---in a restricted way some
correlation among some observations. The introduction of random effects
into growth mixture models attempts to address this need.  A concern,
however, is that random effects also account for a more global type of
correlation, confounding the role of the latent classes and that of the
within-class correlation structure in model identifiability, fitting, and
testing when classes are not crisply separated.  In contrast to random
effects models, auto-regressive processes represent an alternative and
more \emph{local} source of within-subject correlation, allowing observations
close together in time to be more strongly correlated than those further
apart. Local correlation is not at all accounted for by the latent class
structure.  With a within-class local correlation model, the observations
far apart in time will be nearly independent, strengthening model
identifiability.

To address these issues, we propose a longitudinal latent class
model for count data which yields a closed form likelihood, accounts for local
correlation among the repeated measures on a given subject, and allows for
global association to be accounted for by the latent class structure. With our
approach, correlations between observations far apart in time will be especially
informative about class membership because the subject-specific correlation in
these cases will be negligible.

Our contributions in this paper are as follows. In the next section, we
provide a technical description of our discrete data AR-1 process latent
class trajectory model, followed in Section 3 with our approach to
estimating and making inferences on model parameters. There, we rely on a
variation of the EM algorithm that exploits a general estimating function
rather than the true likelihood score function. In Section 4, we propose
a novel measure of the inherent ability of latent class data to
discriminate among classes for subjects in the population. To our
knowledge, this is a previously unexplored, but important construct in
latent class analysis especially when such analysis is used to to assign
subjects to their classes based on their manifest data. Because it is
based on the true data generating mechanism, our measure represents an upper bound on the ability for any fitted statistical model to perform class assignment. We also extend this measure to such fitted models and modeling procedures for finite data.  Section 5 presents a simulation study with the aims of quantifying the statistical operating characteristics of our proposed model in terms of parameter estimation, bias, and confidence interval coverage. We also quantify accuracy of class assignment. Additionally, we examine the ability of our approach to support class assignment when the data generating mechanism is different from the one specified by our model.  Finally, we examine a longitudinal study of conduct problems, and illustrate the use of our model for class definition and assignment in that study.


\section{Model Description}
\label{s:model}
\subsection{Data Structure and Trajectory Model}
\label{ss:structure}
Let $\mathbf{y}_{i} = (y_{i1},\dots,y_{in_{i}})$ be observed longitudinal counts associated with the $i^{th}$ subject.
In total, we have measurements on $m$ subjects $\mathbf{Y} = (\mathbf{y}_{1},\dots,\mathbf{y}_{m})$. We
have observations on subject $i$ at each of the time points $(t_{i1},\dots,t_{in_{i}})$, and we let
$y_{ij}$ denote the observation on subject $i$ at time $t_{ij}$. For each subject and each observed
time point, we also observe a $p \times 1$ covariate vector $\mathbf{x}_{ij}^{T}$, with $\mathbf{X}_{i} = (\mathbf{x}_{i1},
\dots,\mathbf{x}_{in_{i}})^{T}$ denoting the $n_{i} \times p$ design matrix for subject $i$. In
addition, each subject has an unobserved ``latent class'' $Z_{i}$ with $Z_{i} \in \{1,\dots,C\}$
indicating membership in one of $C$ latent classes.

The distribution of $(\mathbf{y}_{i}|\mathbf{X}_{i},Z_{i}=c)$ is governed by a vector of class-specific
parameters $\btheta_{c}$ with $p(\mathbf{y}_{i}|\mathbf{X}_{i},Z_{i}=c) = p_{i}(\mathbf{y}_{i};\btheta_{c})$
denoting the distribution of $\mathbf{y}_{i}$ given covariates $\mathbf{X}_{i}$ and class label $c$. Observations
made on different subjects are assumed to be independent, and the class labels $(Z_{1},\dots,Z_{n})$ are assumed to be i.i.d.\ random
variables with the vector of mixture proportions $\bpi = (\pi_{1},\dots,\pi_{C})$ determining
the class-membership proportions (i.e., $P(Z_{i}=c) = \pi_{c}$) in the population.

Conditional on a subject's class, the mean-response curve or latent trajectory is
\begin{equation}
E(\mathbf{y}_{i}|Z_{i}=c,\mathbf{X}_{i}) = E_{\theta_{c}}(\mathbf{y}_{i}) = \bmu_{i}^{c}
= (\mu_{i1}^{c},\dots,\mu_{in_{i}}^{c})\ ,
\label{eq:meandef}
\end{equation}
where $E_{\theta_{c}}( \cdot )$ denotes taking expectation
conditional on subject $i$ belonging to class~$c$.

We relate the mean curve $\bmu_{i}^{c}$ to the covariates through $\log(\mu_{ij}^{c}) = \mathbf{x}_{ij}^{T}\bbeta_{c}$, where
$\bbeta_{c} = (\beta_{1}^{c},\dots,\beta_{p}^{c})$ are the class-$c$ regression coefficients. To allow
for overdispersion, the variance function has the form $\var_{\theta_{c}}(y_{ij}) = \phi_{c}\mu_{ij}^{c}$,
with scale parameter $\phi_{c}$ allowed to vary across classes. Due to our data-generating model
(Section \ref{ss:inardescription}), we must have $\phi_{c} > 1$. For this reason,
we often write the scale parameter as $\phi_{c} = 1 + \gamma_{c}$, where $\gamma_{c} > 0$.

\subsection{The INAR-(1) Negative Binomial Process}
\label{ss:inardescription}
Conditional on class-membership, the observations from subject $i$ comprise a multivariate outcome with distribution
$p_{i}(\mathbf{y}_{i};\btheta_{c})$, governed by the $(p+2) \times 1$ class-specific parameter vector
$\btheta_{c} = (\bbeta_{c}^{T},\alpha_{c},\phi_{c})^{T}$. The distribution of
$\mathbf{y}_{i} = (y_{i1},\dots,y_{in_{i}})$ is modeled by assuming that the components $y_{ij}$ of $\mathbf{y}_{i}$ arise from
a first-order Markov process governed by $\btheta_{c}$. The joint distribution of $\mathbf{y}_{i}$ is built up directly through the
transition function, $p(y_{ij}|y_{i,j-1},\mathbf{X}_{i};\btheta_{c})$, associated with the underlying process
$p_{i}(\mathbf{y}_{i};\btheta_{c}) = p(y_{i1}|\mathbf{X}_{i};\btheta_{c})\prod_{j=2}^{n_{i}}{ p(y_{ij}|y_{i,j-1},\mathbf{X}_{i};\btheta_{c}) }$. The correlation structure of $\mathbf{y}_{i}$ then arises from the various dependencies introduced 
by the Markov process.

A stochastic process tailored specifically for count data is the integer-valued autoregressive (INAR(1)-NB)
process with negative binomial marginals described in \cite{McKenzie:1986} and \cite{Bockenholt:1999}. For a subject in class $c$, observations
from the INAR(1)-NB process arise as follows: the first observation $y_{i1}$ follows a negative binomial distribution
with $E_{\theta_{c}}(y_{i1}) = \mu_{i1}^{c}$ and $\var_{\theta_{c}}(y_{i1}) = \mu_{i1}^{c}(1 + \gamma_{c})$.
We denote this by $y_{i1} \sim NB\big\{\mu_{i1}^{c},\mu_{i1}^{c}(1 + \gamma_{c})\big\}$ meaning that $y_{i1}$
has probability mass function 
\begin{equation}
P(y_{i1}=k) = {k + \mu_{i1}^{c}/\gamma_{c} - 1 \choose k}
\Big( \frac{1}{1 + \gamma_{c}} \Big)^{\mu_{i1}^{c}/\gamma_{c}} \Big( \frac{\gamma_{c}}{1 + \gamma_{c}} \Big)^{k}; \quad k \geq 0.
\label{eq:negbin_pmf}
\end{equation}
The subsequent values $(y_{i2},\dots,y_{in_{i}})$ are determined through
\begin{equation}
y_{ij} = H_{ij} + I_{ij}, \quad j=2,\ldots,n_{i},  \label{eq:recursion}
\end{equation}
where conditional on $(y_{i1},\ldots,y_{i,j-1})$ and latent probabilities $(q_{i2},\ldots,q_{in_{i}})$, \\
$H_{ij} \sim \textrm{Binomial}(y_{i,j-1},q_{ij})$, with the understanding
that $H_{ij} = 0$ whenever $y_{i,j-1} = 0$. The success probabilities
$(q_{i2},\dots,q_{in_{i}})$ are themselves independent random variables
with
$q_{ij} \sim \textrm{Beta} \big\{ \alpha_{c}\mu_{i,j-1}^{c}/\gamma_{c}, (1 -
\alpha_{c})\mu_{i,j-1}^{c}/\gamma_{c} \big\}$,
where $\textrm{Beta}(\alpha,\beta)$ represents a Beta distribution with
shape parameters $\alpha$ and $\beta$.  Because this implies that that
$E[H_{ijh}|y_{i,j-1}] = \alpha_{c}y_{i,j-1}$ marginally over $q_{ij}$, we
must have $\alpha_{c} \in [0,1]$ for each class. One may also note that
given $y_{i,j-1} \geq 1$ and class membership $c$, $H_{ij}$ follows a
beta-binomial distribution with parameters
$(y_{i,j-1},\alpha_{c}\mu_{i,j-1}^{c}/\gamma_{c}, (1 -
\alpha_{c})\mu_{i,j-1}^{c}/\gamma_{c})$.
The innovation component $I_{ij}$ is assumed to follow a
$\textrm{NB}\big\{ \mu_{ij}^{c}(1 - \alpha_{c}),\mu_{ij}^{c}(1 - \alpha_{c})(1 + \gamma_{c}) \big\}$
distribution where $(I_{i2},\ldots,I_{in_{i}})$ are mutually independent
and where each $I_{ij}$ is independent of the history $(y_{i1},\ldots,y_{i,j-1})$.

Although the transition function $p(y_{ij}|y_{i,j-1},\mathbf{X}_{i},\btheta_{c})$ associated with 
the INAR(1)-NB process does not have a simple closed form (see \cite{Bockenholt:1999}), it
may be directly computed by using the fact that it is the convolution of a beta-binomial distribution and a negative binomial distribution.
In addition, under our description of the INAR(1)-NB process, the marginal
distribution (conditional on class membership) of $y_{ij}$ is negative
binomial with $E_{\theta_{c}}(y_{ij}) = \mu_{ij}^{c}$ and
$\var_{\theta_{c}}(y_{ij}) = \mu_{ij}^{c}(1 + \gamma_{c})$, and the
within-class correlation structure of $\mathbf{y}_{i}$ is first-order
autoregressive. That is, for two observations $y_{ik}$ and $y_{ij}$ on the
same subject, the within-class correlation is
$\corr_{\theta_{c}}(y_{ik},y_{ij}) = \alpha_{c}^{|k-j|}$. The conditional
expectation of $y_{ij}$ given $y_{i,j-1}$ is a linear function of
$y_{i,j-1},$
\begin{equation}
E_{\theta_{c}}(y_{ij}|y_{i,j-1}) = \mu_{ij}^{c}\Big( 1 - \lambda_{ij}^{c} \Big) + \lambda_{ij}^{c}y_{i,j-1},
\label{eq:condexp}
\end{equation}
and the conditional variance of $y_{ij}$ given $y_{i,j-1}$ is given by
\begin{equation}
\var_{\theta_{c}}(y_{ij}|y_{i,j-1}) = \mu_{ij}^{c}\Big( 1 - \lambda_{ij}^{c} \Big)\phi_{c}
+ y_{i,j-1}\lambda_{ij}^{c}\Big( 1 - \lambda_{ij}^{c} \Big)
\frac{\mu_{i,j-1}^{c}/\gamma_{c} + y_{i,j-1}}{1 + \mu_{i,j-1}^{c}/\gamma_{c}},
\label{eq:condvar}
\end{equation}
where $\lambda_{ij}^{c} = \alpha_{c}\sqrt{\mu_{ij}^{c}/\mu_{i,j-1}^{c}}$.

It is also worth mentioning that our specification of the INAR(1)-NB process places the following restriction
on the relation between the autocorrelation parameters and the latent trajectories:
$\alpha_{c}^{2} < \min_{i,j}\{ \mu_{i,j-1}^{c}/\mu_{ij}^{c},\mu_{ij}^{c}/\mu_{i,j-1}^{c} \}$ 
for each $c$.
However, when all of the latent trajectories are reasonably smooth, this constraint is not especially
restrictive as the values of $\{ \mu_{i,j-1}^{c}/\mu_{ij}^{c}  \}$ will be close to one.

\section{Estimation}
\label{s:estimation}
Because a finite mixture model with a fixed number of components can easily be formulated as a ``missing-data''
problem, the EM algorithm provides an attractive estimation approach. In our model, if the individual
class-membership labels $\mathbf{Z} = (Z_{1},\dots,Z_{n})$ were observed, the ``complete-data'' log-likelihood would be
\begin{equation}
\log L(\mathbf{\Theta},\bpi;\mathbf{Y},\mathbf{Z})
= \sum_{i=1}^{m}{ \sum_{c=1}^{C}{ \mathbf{1}\{ Z_{i}=c \} \Big( \log(\pi_{c}) + \log\{ p_{i}(\mathbf{y}_{i};\btheta_{c}) \}   \Big)    }    }.
\label{eq:comp_loglik}
\end{equation}
Above, $\mathbf{\Theta} = (\btheta_{1},\dots,\btheta_{C})$ where $\btheta_{c} = (\bbeta_{c},
\alpha_{c},\gamma_{c})$ are the parameters associated with class $c$, and $\bpi = (\pi_{1},\dots,\pi_{C})$ is
the vector of mixture proportions.

Given current, interation-$k$, estimates of parameters $(\mathbf{\Theta}^{(k)},\bpi^{(k)})$, each EM iteration obtains new
parameter estimates by maximizing the current expectation of the complete-data log-likelihood,
with the expectation being taken over the unobserved class labels, viz.
\begin{eqnarray}
(\mathbf{\Theta}^{(k+1)},\bpi^{(k+1)}) &=& \underset{\Theta,\pi}{\operatorname{argmax }}
E\Big\{  \log L(\mathbf{\Theta},\bpi;\mathbf{Y},\mathbf{Z}) \Big| \mathbf{Y},\mathbf{\Theta}^{(k)},\bpi^{(k)}   \Big\} \nonumber \\
&=& \underset{\Theta,\pi}{\operatorname{argmax }}\sum_{c=1}^{C}{\sum_{i=1}^{m}{
W_{ic}(\mathbf{\Theta}^{(k)},\bpi^{(k)})\Big( \log(\pi_{c}) + \log\{p_{i}(\mathbf{y}_{i};\btheta_{c}\}  \Big)   }}.
\label{eq:emupdate}
\end{eqnarray}
Here, $W_{ic}(\mathbf{\Theta}^{(k)},\bpi^{(k)})$ is the current estimated posterior probability
that subject $i$ belongs to class $c$, namely
\begin{equation}
W_{ic}(\mathbf{\Theta}^{(k)},\bpi^{(k)}) = P(Z_{i}=c \big| \mathbf{y}_{i},\mathbf{\Theta}^{(k)},\bpi^{(k)})
= \frac{ \pi_{c}p_{i}(\mathbf{y}_{i};\btheta_{c}^{(k)}) }{\sum_{s}{\pi_{s}p_{i}(\mathbf{y}_{i};\btheta_{s}^{(k)})}  }.
\label{eq:postprobs}
\end{equation}
\subsection{Estimating Equation Approach for Computation}
\label{ss:eeapproach}
In each step of the EM algorithm, updating the class probabilities is straightforward because the update is simply the
average of the current posterior probabilities,
$\pi_{c}^{(k+1)} = \frac{1}{m}\sum_{i=1}^{m}{ W_{ic}^{(k)}(\bTheta^{(k)},\bpi^{(k)}) }$.
However, to update the remaining parameters, we must maximize $K$ separate weighted log-likelihood functions
\begin{equation}
\btheta_{c}^{(k+1)} = \underset{\theta_{c}}{\operatorname{argmax }}
\sum_{i=1}^{m}{ W_{ic}(\bTheta^{(k)},\bpi^{(k)})\log\{p_{i}(\mathbf{y}_{i};\btheta_{c}) \} }, \qquad c=1,\ldots,C.
\label{eq:weightedlogliks}
\end{equation}
Because each such log-likelihood function is a sum over many complicated transition probabilities,
implementing the maximization in (\ref{eq:weightedlogliks}) may be challenging.

Instead of updating the parameters by maximizing each of the weighted log-likelihood functions directly,
we found that replacing the score function with a more manageable estimating function provides
a less cumbersome approach. That is, letting $s_{i}(\btheta_{c}) = \partial \log p_{i}(
\mathbf{y}_{i};\btheta_{c})/\partial \btheta_{c}$ denote the score function, we suggest that, rather
than solving
\begin{equation}
\sum_{i=1}^{m}{ W_{ic}(\mathbf{\Theta}^{(k)},\bpi^{(k)})s_{i}(\btheta_{c}) } = \mathbf{0}
\end{equation}
for each class, one instead solve
\begin{equation}
\sum_{i=1}^{m}{ W_{ic}(\bTheta^{(k)},\bpi^{(k)})U_{i}(\btheta_{c})  } = \mathbf{0};
\qquad \textrm{for } c=1,\ldots,C,
\label{eq:weightedgee}
\end{equation}
where $ U_{i}( \btheta_{c} )$ forms an unbiased
estimating function (i.e., $E_{\theta_{c}}[ U_{i}(\btheta_{c}) ] = \mathbf{0}$) for each class.
Such an approach, where within each EM iteration the maximization step is approximated by solving an estimating equation,
is similar to the estimation strategy described in \cite{Elashoff:2004}.

Our choice of $U_{i}(\btheta_{c})$ relies on the extended quasilikelihood
function procedure for constructing estimating equations that was proposed
in \citet{Hall:1998}. These estimating functions considerably simplify
computation (i.e., by solving (\ref{eq:weightedgee})) when compared to
using score functions $s_{i}( \btheta_{c} )$. Tailoring \citet{Hall:1998} to the case of a log-link function and an
AR(1) correlation structure yields the $(p+2) \times 1$ estimating
function
\begin{equation}
U_{i}(\btheta_{c}) =
\left[ \begin{array}{cc} U_{i}^{[1]}(\btheta_{c}) \\
U_{i}^{[2]}(\btheta_{c}) \\  U_{i}^{[3]}(\btheta_{c}) \end{array} \right]
= \frac{1}{\phi_{c}}\left[ \begin{array}{cc}
\mathbf{X}_{i}^{T}\mathbf{A}_{i}^{1/2}(\bmu_{i}^{c})\mathbf{R}_{i}^{-1}(\alpha_{c})\mathbf{A}_{i}^{-1/2}(\bmu_{i}^{c})(\mathbf{y}_{i} - \bmu_{i}^{c}) \\
\frac{2\phi_{c}\alpha_{c}(n_{i} - 1)}{1 - \alpha_{c}^{2}} - (\mathbf{y}_{i} - \bmu_{i}^{c})^{T}\frac{d\mathbf{R}_{i}^{-1}(\alpha_{c})}{d\alpha_{c}}
(\mathbf{y}_{i} - \bmu_{i}^{c}) \\
\frac{1}{\phi_{c}}(\mathbf{y}_{i} - \bmu_{i}^{c})^{T}\mathbf{A}_{i}^{-1/2}(\bmu_{i}^{c})\mathbf{R}_{i}^{-1}(\alpha_{c})\mathbf{A}_{i}^{-1/2}(\bmu_{i}^{c})
(\mathbf{y}_{i} - \bmu_{i}^{c}) - n_{i}
\end{array} \right]\ .
\label{eq:maingee}
\end{equation}
In (\ref{eq:maingee}), $\mathbf{A}_{i}(\bmu_{i}^{c})$ is the $n_{i} \times n_{i}$ matrix defined by $A_{i}(\bmu_{i}^{c})
= \textrm{diag}\{\mu_{i1},\dots,\mu_{in_{i}}\}$
and $\mathbf{R}_{i}(\alpha_{c})$ is the $n_{i} \times n_{i}$ correlation matrix
whose $(k,j)$ entry is $R_{i}(\alpha_{c})[k,j] = \alpha_{c}^{|k-j|}$.

The equation determined by setting the $p$-component vector $\sum_{i=1}^{m}{ U_{i}^{[1]}(\btheta_{c}) }$ to zero
\begin{equation}
\frac{1}{\phi_{c}}\sum_{i=1}^{m}{\mathbf{X}_{i}^{T}\mathbf{A}_{i}^{1/2}(\bmu_{i}^{c})\mathbf{R}_{i}^{-1}(\alpha_{c})\mathbf{A}_{i}^{-1/2}(\bmu_{i}^{c})
(\mathbf{y_{i}} - \bmu_{i}^{c}) } = \mathbf{0}, \label{eq:gee}
\end{equation}
corresponds to the generalized estimating equation (GEE) described in \citet{Zeger1986}.
In GEE, the autocorrelation parameter $\alpha_{c}$ is first estimated separately and then plugged into (\ref{eq:gee})
in order to solve for regression coefficients~$\bbeta_{c}$.
In contrast, solving (\ref{eq:weightedgee}) requires that $\alpha_{c}$ be
estimated simultaneously with~$\bbeta_{c}$.
To solve (\ref{eq:weightedgee}) for a fixed $c$, we first update $\bbeta_{c}$ by solving 
$\sum_{i=1}^{m}{W_{i,c}(\bTheta^{(k)},\bpi^{(k)})U_{i}^{[1]}(\btheta_{c})} = \mathbf{0}$, using initial values for $(\alpha_{c},\phi_{c})$.
Using this value of $\bbeta_{c}$ and the initial overdispersion $\phi_{c}$, $\alpha_{c}$ is updated by solving 
$\sum_{i=1}^{m}{W_{i,c}(\bTheta^{(k)},\bpi^{(k)})U_{i}^{[2]}(\btheta_{c})} = 0$.
The value of $\phi_{c}$ can then be updated non-iteratively because, given values of $(\bbeta_{c},\alpha_{c})$,
solving $\sum_{i=1}^{m}{W_{i,c}(\bTheta^{(k)},\bpi^{(k)})U_{i}^{[3]}(\btheta_{c})} = 0$ for $\phi_{c}$ has a closed form.
This procedure is repeated until convergence.

\subsection{Quasi-EM Procedure and Standard Errors}
\label{ss:stderrs}
Our altered EM algorithm, where the score function is replaced with another, more manageable estimating function,
may be summarized as follows:
\begin{enumerate}
\item[\textbf{1.}]
Find initial estimates $\bTheta^{(0)}$, $\bpi^{(0)}$. (Our initialization 
procedure is described in the appendix).
\item[\textbf{2.}]
Compute current estimated posterior probabilities $W_{ic}(\bTheta^{(k)},\bpi^{(k)})$ for each class and
subject.
\item[\textbf{3.}]
Update mixture proportions through $\pi_{c}^{(k+1)} = \frac{1}{m}\sum_{i=1}^{m}{ W_{ic}(\bTheta^{(k)},\bpi^{(k)}) }$.
Update other parameters $(\btheta_{1},\dots,\btheta_{K})$ by solving
$\sum_{i=1}^{m}{W_{ic}(\bTheta^{(k)},\bpi^{(k)}) U_{i}(\btheta_{c})} = \mathbf{0}$, for $c=1,\dots,C$.
\item[\textbf{4.}]
Repeat steps (2) and (3) until convergence.
\end{enumerate}
Parameter estimates $(\widehat{\bTheta},\hat{\bpi})$ produced from the above iterative procedure may 
be viewed as a solution to the estimating equation
$G(\bTheta,\bpi) = \sum_{i=1}^{m}{G_{i}(\bTheta,\bpi)} = \mathbf{0}$, where
$G_{i}(\bTheta,\bpi)$ is the $(p+3)C - 1 \times 1$ vector defined as
$G_{i}(\bTheta,\bpi) = [\textrm{vec}(\mathbf{V}_{i}),\mathbf{b}_{i}]^{T}$ and where
$\mathbf{V}_{i}$ is the $(p+2) \times C$ matrix $\mathbf{V}_{i} = \big[
W_{i1}(\bTheta,\bpi)U_{i}(\btheta_{1}), \ldots\ldots ,W_{iC}(\bTheta,\bpi)U_{i}(\btheta_{C}) \big]$
and $\mathbf{b}_{i}$ is the $(C-1) \times 1$ vector $\mathbf{b}_{i} = \big[
W_{i1}(\bTheta,\bpi) - \pi_{1}, \ldots\ldots, W_{i,C-1}(\bTheta,\bpi) - \pi_{K-1}  \big]^{T}$.

If we let $U_{i}^{k}(\btheta_{c};\mathbf{y_{i}})$ be the $k^{th}$ component of $U_{i}(\btheta_{c})$,
we can see that $G(\bTheta,\bpi) = \mathbf{0}$ forms an unbiased estimating equation for
$(\bTheta,\bpi)$ because the expectation of the $(k,c)$ component of $\mathbf{V}_{i}$
is given by
\begin{equation}
E\Big\{ W_{ic}(\bTheta,\bpi)U_{i}^{k}(\btheta_{c};\mathbf{y}_{i}) \Big\}
= \pi_{c}E\Big\{ \frac{p_{\theta_{c}}(\mathbf{y}_{i})}{p(\mathbf{y}_{i})}
U_{i}^{k}(\btheta_{c};\mathbf{y}_{i}) \Big\} = \pi_{c}E_{\theta_{c}}\Big\{ U_{i}^{k}(\btheta_{c};\mathbf{y}_{i})
\Big\} = 0\ ,
\end{equation}
and the expectation of the $c^{th}$ element of $\mathbf{b}_{i}$ is given by
\begin{equation}
E\Big\{ W_{ic}(\mathbf{\Theta},\bpi) - \pi_{c} \Big\} 
= \pi_{c}E\Big\{ \frac{p_{\theta_{c}}(\mathbf{y_{i}})}{p(\mathbf{y_{i}})} \Big\} - \pi_{c} = 0.
\end{equation}
Conditions under which solutions of unbiased estimating equations are asymptotically normal
(after normalizing appropriately) are discussed in a number of sources (see e.g., section 12.4
of \cite{Heyde:1997}, or \cite{Crowder:1986}). By applying the typical results regarding
asymptotic normality to our setting, we have that
$\widehat{\bSigma}^{-1/2}\big\{ (\widehat{\bTheta},\hat{\mathbf{\bpi}})^{T} - (\bTheta,\bpi)^{T} \big\}
\longrightarrow_{d} N(\mathbf{0},\mathbf{I}_{qK-1})$ as $m \longrightarrow \infty$, where
$\widehat{\bSigma}$ is given by
\begin{equation}
\widehat{\bSigma} = \Big( \frac{1}{m} \sum_{i=1}^{m}{ E\big\{ \mathbf{DG}_{i}(\widehat{\bTheta},\hat{\bpi}) \big\} } \Big)^{-1}
\Big( \frac{1}{m}\sum_{i=1}^{m}{\mathbf{G}_{i}(\widehat{\bTheta},\hat{\bpi})\mathbf{G}_{i}(\widehat{\bTheta},\hat{\bpi})^{T}} \Big)
\Big( \frac{1}{m}\sum_{i=1}^{m}{ E\big\{ \mathbf{DG}_{i}(\widehat{\bTheta},\hat{\bpi}) \big\} } \Big)^{-1^{T}}.
\label{eq:asympcovmat}
\end{equation}
In (\ref{eq:asympcovmat}), $\mathbf{DG}_{i}(\bTheta,\bpi)$ is the $\big( (p+3)C - 1 \big) \times \big( (p+3)C - 1 \big)$ 
matrix of partial derivatives $\mathbf{DG}_{i}(\btheta,\bpi) = \partial \mathbf{G}_{i}(\bTheta,\bpi)/\partial (\bTheta,\bpi)$.
We compute standard errors using the diagonal elements of $\widehat{\bSigma}$.

\section{Class Assignment and Discrimination Measures}
\label{s:sims}
\subsection{Class Separation Index and Expected Empirical Discrimination}
\label{ss:csi}
After estimating latent trajectories for each class, one often wishes to
go back and assign or classify subjects based on their empirical
resemblance to one of the estimated trajectories.  Even though all the
model parameters may be estimated accurately, however, the associated
classification procedure may not have a high rate of correct
assignment. Poor class-discrimination may be due to little underlying
class-separation between the latent trajectories $\bmu_{i}^{c}$ or to high
levels of noise $\phi_{c}$ in the response.  In this case, the ability to
correctly classify subjects is limited by the class separation inherent in
the underlying true data generating model. To quantify this inherent class
separation, we propose a novel class separation index (CSI) which provides
a rough upper bound on one's ability to correctly assign subjects to
classes based on our or any other latent class model. To accomplish this,
the CSI measures the classification performance that would be attained if
the underlying true generative model were known.

To fix notation, consider data of the form $\mathbf{Y} = (\mathbf{y}_{1},\ldots,\mathbf{y}_{m})$ and 
let \\ $A_{C}(\mathbf{Y}) = [\mathbf{a}_{1}(\mathbf{Y}), \ldots ,\mathbf{a}_{m}(\mathbf{Y}) ]^{T}$ 
be a procedure which maps data $\mathbf{Y}$ into an $m \times C$ matrix 
of nonnegative entries whose rows sum to one; the $(i,c)$ entry of this matrix can be thought
of as a reported probability that subject $i$ belongs to class $c$. For instance,
we could have $A_{C}(\mathbf{Y}) = \mathbf{W}^{C}(\bTheta,\bpi;\mathbf{Y})$, where 
$\mathbf{W}^{C}(\bTheta,\bpi;\mathbf{Y}) = [\mathbf{w}^{C}(\bTheta,\bpi;\mathbf{y}_{1}), \ldots, 
\mathbf{w}^{C}(\bTheta,\bpi;\mathbf{y}_{m}) ]^{T}$
denotes the $m \times C$ matrix whose $i^{th}$ row, $\mathbf{w}^{C}(\bTheta,\bpi;\mathbf{y}_{i})^{T}$, 
contains the posterior probabilities for subject $i$ computed with
the parameters $(\bTheta,\bpi)$. Alternatively, we could have $A_{C}(\mathbf{Y}) 
= V(\boeta,\mathbf{Y})$, where $V(\boeta,\mathbf{Y})$
denotes a matrix of class membership probabilities computed under an
incorrect working model with specified parameter $\boeta$. Or, 
we could have $A_{C}(\mathbf{Y}) 
= V(\widehat{\boeta}(\mathbf{Y}),\mathbf{Y})$, where
$\widehat{\boeta}(\mathbf{Y})$ is an estimate of $\boeta$ computed with
the available data~$\mathbf{Y}$.

To measure how well a procedure $A_{C}(\mathbf{Y})$ predicts class membership, we will
use a discrimination index $D$, which, given class labels $\mathbf{Z} = (Z_{1},\ldots,Z_{m})$, 
returns a score in $[0,1]$ such that
$D\big( \mathbf{Z},A_{C}(\mathbf{Y}) \big) = 1$ if $A_{C}(\mathbf{Y})$ has
a $1$ in the correct cell for all observations (rows), 
and is less than or equal to one otherwise, and for any $D(\cdot,\cdot)$ considered, 
values closer to one will imply better classification performance. For instance, 
$D(\cdot,\cdot)$ often has a form similar to a $U$-statistic of order $C$ with kernel 
$h$, namely
\begin{equation}
D\big( \mathbf{Z}, A_{C}(\mathbf{Y}) \big) = \frac{1}{N_{1}\cdots N_{C}}
\sum_{i_{1},...,i_{C}}{ h\Bigg(  \colvec{ Z_{i_{1}} \\ \mathbf{a}_{i_{1}}(\mathbf{Y}) },\ldots, 
\colvec{ Z_{i_{C}} \\ \mathbf{a}_{i_{C}}(\mathbf{Y}) } \Bigg) }, 
\label{eq:ustatistic}
\end{equation}
where $N_{c} = \sum_{i=1}^{m}{ \mathbf{1}\{ Z_{i} = c \} }$ and where the summation 
is performed over all values of $(i_{1},\ldots,i_{C})$ 
such that no two members of $(i_{1},\ldots,i_{C})$ are the same and each $1 \leq i_{j} \leq m$
which means that the summation in (\ref{eq:ustatistic}) is performed over ${m \choose C}C!$ terms.
Examples of $D(\cdot,\cdot)$ such as the $c$-statistic and the polytomous discrimination 
index are discussed in Section \ref{ss:dismeasures}.

The class separation index may be defined for any latent class model having responses
$\mathbf{y}_{i}$, latent class indicators $Z_{i}$ and whose data generating mechanism
is governed by the parameters $(\bTheta, \bpi)$. In particular, for a choice of 
index $D(\cdot,\cdot)$, the CSI is defined as
\begin{equation}
CSI = \lim_{m \longrightarrow \infty}E\Big\{ D\big( \mathbf{Z}, \mathbf{W}^{C}(\bTheta,\bpi;\mathbf{Y}) \big) \Big\},
\label{eq:csi}
\end{equation}
provided that the above limit exists.
Heuristically, the CSI is simply the expected discrimination that would be obtained in the long run
by an oracle using knowledge of both the true data-generating model and the true parameter values
to compute the posterior probabilities of class membership. 
Hence, the CSI is a population quantity measuring the underlying separation of the classes and depends on 
neither the sample size nor on the particular way the data have been analyzed.
For example, with an index $D(\cdot,\cdot)$ of the form (\ref{eq:ustatistic}), the 
CSI would be
\begin{equation}
CSI = \frac{1}{\pi_{1}\cdots\pi_{C}} E\Bigg\{ h \Bigg( \colvec{ Z_{1} \\ \mathbf{w}^{C}(\bTheta,\bpi; \mathbf{y}_{1}) },\ldots, 
\colvec{ Z_{r} \\ \mathbf{w}^{C}(\bTheta,\bpi; \mathbf{y}_{C}) } \Bigg) \Bigg\}.
\label{eq:csi_ustat}
\end{equation}

Turning to finite samples, the realized or empirical discrimination
resulting from using procedure $A_{C}( \mathbf{Y} )$ 
is $D\big( \mathbf{Z}, A_{C}( \mathbf{Y} ) \big)$, and the 
expectation of this quantity is what we define to be the 
expected empirical discrimination (EED), namely
\begin{equation}
EED = E\Big\{ D\big( \mathbf{Z}, A_{C}(\mathbf{Y}) \big) \Big\},
\label{eq:eed}
\end{equation}
where the expectation in (\ref{eq:eed}) is taken over
$(\mathbf{Z},\mathbf{Y})$ for a given sample size $m$, under the true data generating model. Note that
the EED may be computed for any procedure $A_{C}( \mathbf{Y} )$ even when, for example,
$A_{C}( \mathbf{Y} )$ corresponds to randomly
generated predictions or contains predicted probabilities computed under 
an incorrect working model. Hence, comparing
the CSI and the EED may serve as a robustness check in the sense that
the CSI may be calculated for any number of hypothetical alternative
models, and for each of these, the EED may be computed using $A_{C}(\mathbf{Y})$ from the assumed working model.
For example, we could compute the CSI under a model with Poisson
responses and Normal random effects using (\ref{eq:csi}) and 
compare it with the EED obtained (taking expectation under the Poisson-Normal model) by 
using our INAR(1)-NB model to find posterior probabilities of class membership.
We explore such a comparison with a small simulation study in Section~5.2.

\subsection{Measures of Discrimination for Multiple Classes}
\label{ss:dismeasures}
When predicting a binary outcome or class label, a widely used measure of a model's ability to discriminate
between the two class labels is the concordance index or $c$-statistic.
If we let $Z_{i} \in \{1,2\}$ denote the class labels and let $\hat{p}_{i}(c)$ denote the
predicted probability that $Z_{i}=c$, $c=1,2$, then the $c$-statistic $C_{12}$ is defined as
\begin{equation}
C_{12}(\mathbf{Z},\hat{\mathbf{p}}(1),\hat{\mathbf{p}}(2)) 
= \frac{1}{N_{1}N_{2}}\sum_{i \in A_{1}}{ \sum_{j \in A_{2}}{ \mathbf{1}\{ \hat{p}_{i}(1) > \hat{p}_{j}(1) \} }  }
+ \frac{1}{2N_{1}N_{2}}\sum_{i \in A_{1}}{ \sum_{j \in A_{2}}{ \mathbf{1}\{ \hat{p}_{i}(1) = \hat{p}_{j}(1) \} }  },
\label{eq:c_statistic}
\end{equation}
where $A_{c} = \{i: Z_{i}=c\}$ is the set of subjects with class label c and
$N_{c} = \sum_{i=1}^{m}{ \mathbf{1}\{ Z_{i} = c \} }$. 
Letting $\hat{\mathbf{p}}_{i} = [\hat{p}_{i}(1),\hat{p}_{i}(2)]^{T}$, 
the $c$-statistic has the form (\ref{eq:ustatistic}) with $C = 2$ and kernel
\begin{equation}
h\Bigg(  \colvec{ Z_{1} \\ \hat{\mathbf{p}}_{1}}, 
\colvec{ Z_{2} \\ \hat{ \mathbf{p} }_{2} } \Bigg)
= \mathbf{1}\{ Z_{1} = 1\}\mathbf{1}\{ Z_{2} = 2\}
\Big( \mathbf{1}\{ \hat{p}_{1}(1) > \hat{p}_{2}(1) \} + \frac{\mathbf{1}\{ \hat{p}_{1}(1) = \hat{p}_{2}(1) \}}{2} \Big).
\end{equation}
When interest lies in assessing the discriminatory capabilities of a classification
method for a multi-class problem, it may be useful to consider multi-class extensions of the
$c$-statistic. One such extension is
the all-pairwise $c$-statistic (see \cite{Hand:2001}) which is simply the average of the 
usual pairwise $c$-statistics (\ref{eq:c_statistic})
taken over all possible pairs of classes. For a $C$-category
outcome, the all-pairwise $c$-statistic ($\textrm{APC}_{C}$) is defined to be
\begin{equation}
\textrm{APC}_{C} = {C\choose 2}^{-1}\sum_{k < j}{ C_{kj}(\mathbf{Z},\hat{\mathbf{p}}(k),\hat{\mathbf{p}}(j))},
\label{eq:pairwise}
\end{equation}
where $C_{kj}(\cdot)$ is defined as in (\ref{eq:c_statistic}) and is computed using only subjects from classes $k$ and~$j$.
Like the $c$-statistic, the all-pairwise $c$-statistic lies between $0$ and $1$
with larger values indicating better discriminatory performance and with values near $0.5$
indicating that the model performs no better than predicting labels
at random.

\begin{figure}
\centering
\includegraphics[width=6in, height=3.5in]{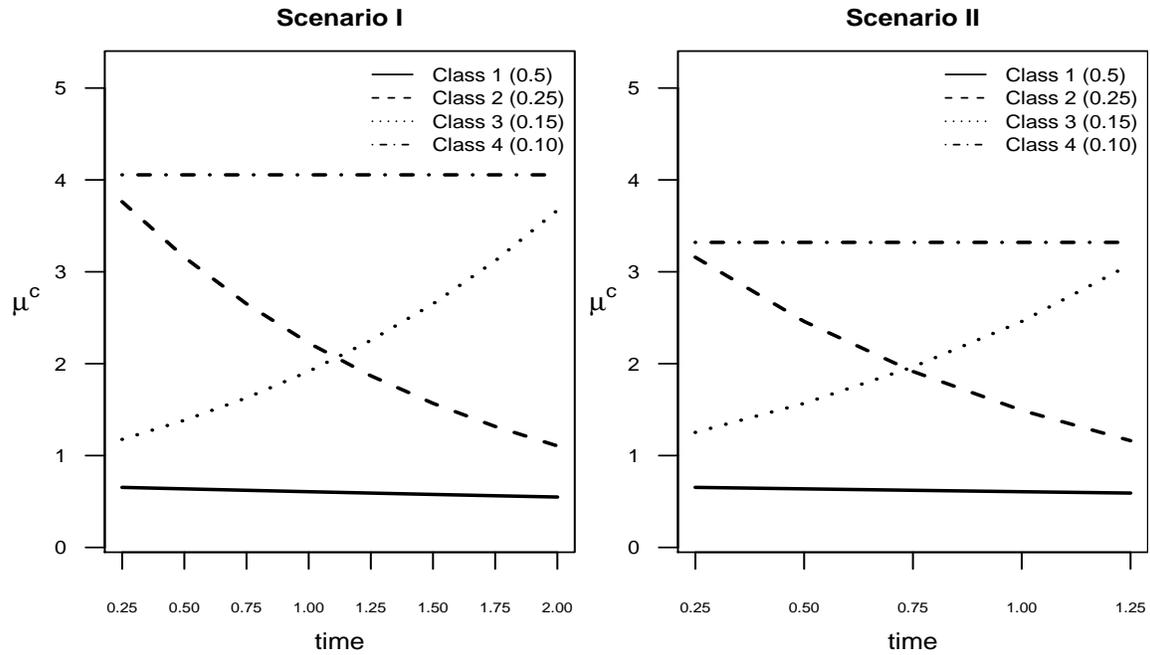}
 \caption{   {\small{Mean latent trajectories $\mu^{c}(t)$ for the two central simulation scenarios. In Scenario I,
 observations for each subject are made at each of the eight time points $t_{ij} = j/4$, $j=1,\ldots,8$,
 and in Scenario II, observations for each subject are made at each of the
 five time points $t_{ij} = j/4$, $j=1,\ldots,5$. For both scenarios, the class-membership
 proportions are as follows: Class 1: $50\%$, Class 2: $25\%$, Class 3: $15\%$, and
 Class 4: $10\%$.}}   }
\label{fig:scenarios}
\end{figure}
Other multi-class discrimination measures include the volume under the ROC surface
and ``one-versus-rest'' measures (see e.g., \cite{Mossman:1999}).
One discrimination index which uses the entire set of predicted probabilities rather than
averaging over the pairwise measures is the Polytomous Discrimination Index (PDI)
proposed in \citet{VanCalster:2012}. Before providing the definition of the PDI,
we first let $\hat{\mathbf{p}}_{i} = (\hat{p}_{i}(1),\ldots,\hat{p}_{i}(C))$
denote the $i^{th}$ subject's vector of predicted probabilities with $\hat{p}_{i}(c)$ representing
the predicted probability that subject $i$ belongs to class $c$.
Then, for a $C$-class model, the PDI is defined to be
\begin{equation}
\textrm{PDI}_{C} = \frac{1}{C N_{1}\cdots N_{C}}
\sum_{i_{1} \in A_{1}}{ \ldots \sum_{i_{C} \in A_{C}}{ 
\sum_{c=1}^{C}{ g_{c}\big( \hat{\mathbf{p}}_{i_{1}},\ldots, \hat{\mathbf{p}}_{i_{C}} \big) } } } ,
\label{eq:multic}
\end{equation}
where $A_{c} = \{i: Z_{i}=c\}$ is the set of subjects in class c and where 
$g_{c}\big( \hat{\mathbf{p}}_{i_{1}},\ldots, \hat{\mathbf{p}}_{i_{C}} \big)$ 
equals one if $\hat{p}_{i_{c}}(c) > \hat{p}_{i_{j}}(c)$ for all $j \neq c$,
and equals zero if there is a $j^{*} \neq c$ such that $\hat{p}_{i_{c}}(c) < \hat{p}_{i_{j^{*}}}(c)$.
If $(\hat{p}_{i_{1}}(c),\ldots,\hat{p}_{i_{C}}(c))$ does not contain a unique maximizer
and $\hat{p}_{i_{c}}(c)$ is one of those tied for the maximum value, then
one sets $g_{c}\big( \hat{\mathbf{p}}_{i_{1}},\ldots, \hat{\mathbf{p}}_{i_{C}} \big) = 1/t$, 
where $t$ is the number of cases tied with $\hat{p}_{i_{c}}(c)$. Unlike the $c$-statistic
or all-pairwise $c$-statistic, a method producing predictions at random will
generate a $\textrm{PDI}_{C}$ value near $1/C$ rather than $0.5$.

\begin{table}[h]
\caption{   {\small{Class separation indices for each of the two central scenarios with
several different values of the autocorrelation and scale parameters. The class separation
indices are computed using both the all-pairwise $c$-statistic ($\textrm{APC}_{C}$)
and the polytomous discrimination index $(\textrm{PDI}_{C})$ as measures of classification performance.}}   }
\begin{center}
{\renewcommand{\arraystretch}{1.2} 
    \begin{tabular}{l| ll|ll | ll|ll   }
     \toprule
      \multicolumn{1}{c}{}& \multicolumn{4}{ c| }{Scenario I} & \multicolumn{4}{ |c }{Scenario II} \\
     \cline{2-9}
     \multicolumn{1}{c}{} & \multicolumn{2}{ c| }{$\phi=1.25$} & \multicolumn{2}{ |c| }{$\phi=3.0$} & \multicolumn{2}{ |c| }{$\phi=1.25$} & \multicolumn{2}{ |c }{$\phi=3.0$} \\
    \cline{2-9}
    \multicolumn{1}{c}{} & $\alpha = 0.1$ & $\alpha = 0.4$ & $\alpha = 0.1$ & $\alpha = 0.4$ & $\alpha = 0.1$ & $\alpha = 0.4$ & $\alpha = 0.1$ & $\alpha = 0.4$ \\
    \hline
    $\textrm{APC}_{C}$ & 0.976 & 0.944 & 0.922 & 0.892 & 0.900 & 0.867 & 0.828 & 0.802  \\
    $\textrm{PDI}_{C}$ & 0.934 & 0.872 & 0.812 & 0.756 & 0.775 & 0.712 & 0.646 & 0.608 \\
    \bottomrule
    \end{tabular}
}
\end{center}
\label{table:csi}
\end{table}

\section{Simulations}
\label{s:simscenarios}
\subsection{Autoregressive Models with Four Latent Classes}
\label{ss:arsims}

\noindent \emph{Design}

To evaluate the performance of our estimation procedure and to test our
implementation, we performed a simulation study using two central
scenarios (Scenarios I and II) as our point of reference.
Each of Scenarios I and II involves a model with four latent classes where
the within-class model is the autoregressive negative binomial model
described in Sections \ref{ss:structure} and \ref{ss:inardescription}. In
Scenario I, each subject is observed over $8$ equally spaced time points,
and in Scenario II, subjects are observed over $5$ time points. As shown
in Figure \ref{fig:scenarios}, the latent trajectories for both of these
scenarios are qualitatively similar. The choice of four classes for the
simulation scenarios is meant to reflect the wide use of four-class models
when identifying subtypes in the childhood development of conduct
disorders (see, e.g., \citet{Odgers:2007} or \cite{Moffitt:1993}).  The
goals of the simulation study include evaluating the classification
performance associated with using the estimated posterior probabilities,
examining the empirical bias of parameter estimates, quantifying the
degree to which the standard errors provide approximately desired coverage
levels, and assessing how each of these operational characteristics vary
as a function of the class separation index.

\begin{figure}
\centering
     \includegraphics[width=6in,height=3in]{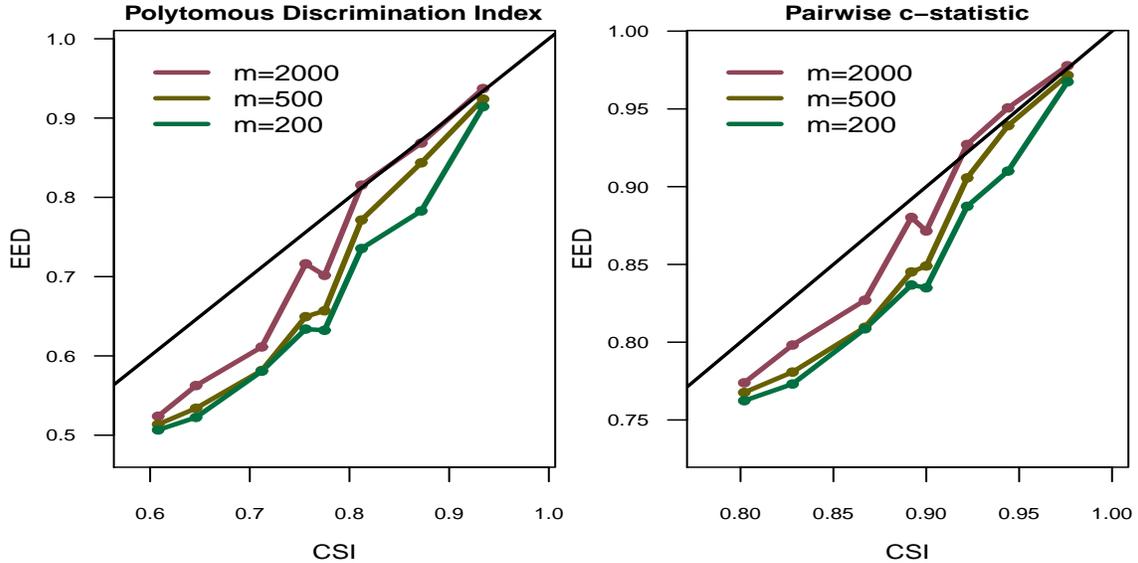}
\caption{    {\small{Expected empirical discrimination (EED) and 
  Class Separation Indices (CSI) for Scenarios I and II of the INAR(1)-NB model.
  Values of the EED are shown when the parameters are estimated from simulations 
  with $m=200$, $m=500$, $m=2,000$ subjects.}}   }
\label{fig:discrimination}
\end{figure}

The values of the class separation index for Scenarios I and II and their
variants are displayed in Table \ref{table:csi}.  For each of the two
scenarios, we varied the autocorrelation parameter across two levels,
$\alpha \in \{ 0.1, 0.4 \},$ and varied the scale parameter across two
levels, $\phi \in \{ 1.25, 3\}.$ As may be inferred from Table
\ref{table:csi}, higher values of the scale parameter and higher levels of
autocorrelation reduce the class separation and thereby the inherent
ability to correctly classify individuals.  Also, for each parameter
configuration (i.e., a particular scenario and choice of $(\alpha,\phi)$),
we ran our estimation procedure for each of the sample sizes $m = 2,000$,
$m = 500$, and $m = 200$.  For each parameter configuration and
sample size, we computed estimates on each of $200$ replicated data
sets. To determine convergence with a given tolerance level $\varepsilon$,
we used the criterion
$\max_{k}|m^{-1}G^{k}(\widehat{\mathbf{\Theta}},\hat{\bpi})| \leq \varepsilon$,
where $G(\mathbf{\Theta},\bpi)$ is as defined in Section \ref{ss:stderrs},
and $G^{k}(\mathbf{\Theta},\bpi)$ denotes the $k^{th}$ element of
$G(\mathbf{\Theta},\bpi)$.

Because the class labels are not identifiable, 
for each model fitted, we permuted the class labels of the estimated
parameters in order to minimize the $L_{2}$ distance between the estimated and true mean curves.
That is, for a set of estimates $\widetilde{\bTheta} = (\tilde{\btheta}_{1},\ldots,\tilde{\btheta}_{4})$ and
$\tilde{\bpi} = (\tilde{\pi}_{1},\ldots,\tilde{\pi}_{4})$ obtained through
our quasi-EM procedure, we computed the optimal permutation according to
\begin{equation}
s^{*} = \underset{S \in \mathcal{P}}{\operatorname{argmin }}\sum_{c=1}^{4}{ (\bmu^{c} - \hat{\bmu}^{S(c)})^{T}(\bmu^{c} - \hat{\bmu}^{S(c)}) },
\end{equation}
where $\mathcal{P}$ simply denotes the set of permutations of class labels $(1,2,3,4)$. We then
computed the final estimates of the parameters through
$\widehat{\bTheta} = (\tilde{\btheta}_{s^{*}(1)},\ldots,\tilde{\btheta}_{s^{*}(4)})$ and
$\hat{\bpi} = (\tilde{\pi}_{s^{*}(1)},\ldots,\tilde{\pi}_{s^{*}(4)})$. Note that
$\bmu^{c} = \exp(\mathbf{X}_{i}\bbeta_{c})$ and $\hat{\bmu}^{c} = \exp(\mathbf{X}_{i}\widehat{\bbeta}_{c})$
do not depend on $i$ since all subjects share the same design matrix in our simulation scenarios,
though this need not be the case in real applications.

\medskip

\noindent \emph{Results}

Figure \ref{fig:discrimination} displays the discriminatory performance (using both the polytomous discrimination index
and the all-pairwise $c$-statistic) of our estimation procedure across the eight simulation settings.
For settings with highly distinct classes, the EED obtained from 
using the estimated posterior probabilities compares favorably with the oracle procedure
even for relatively modest sample sizes (i.e., $m = 200$ and $m = 500$) though the gap between
the EED and the oracle-based CSI tends to widen noticeably as the classes
become less distinguishable. Notably, the $m=2,000$ simulations illustrate that, for highly distinct
classes and reasonably large sample sizes, the discriminatory performance of our procedure
is nearly identical to that of the oracle procedure.  

\begin{figure}
\centering
     \includegraphics[width=4.5in,height=3.5in]{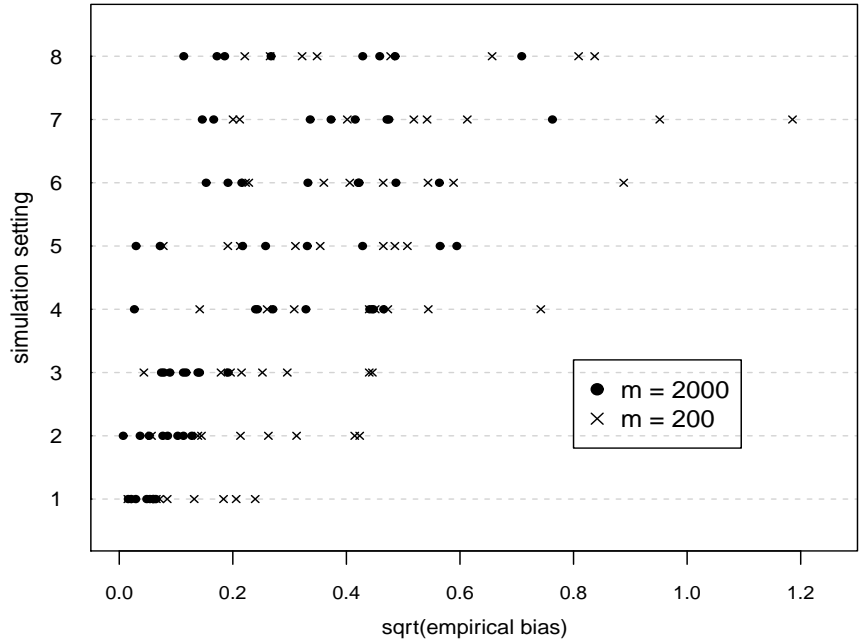}
     \caption{ {\small{Absolute empirical bias for the regression
           coefficients using data simulated under eight settings of the
           INAR(1)-NB model (i.e., each of the scenarios from Table
           \ref{table:csi}).  The simulation settings are ordered
           according to the class separation index with simulation setting $1$
           having the highest class separation and simulation setting $8$
           having the lowest class separation.  The values of the
           empirical bias were obtained using $200$ replications for each
           of the eight scenarios and each choice of the number of
           subjects (either $m = 200$ or $m = 2,000$).}}  }
\label{fig:bias_plot}
\end{figure}

\begin{figure}
\centering
     \includegraphics[width=5.5in,height=4in]{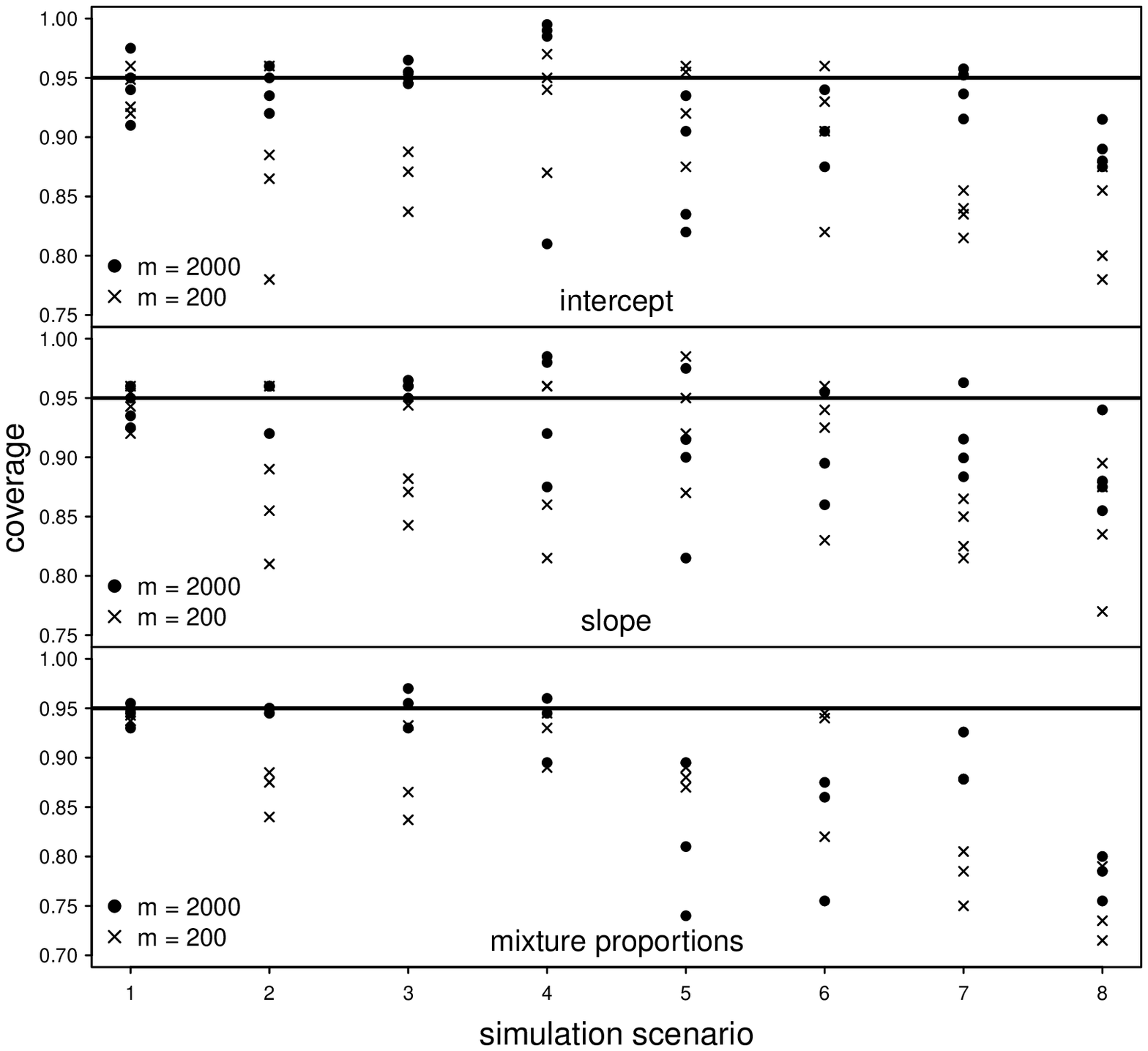}
\caption{    {\small{Coverage proportions (using $95\%$ confidence intervals) for the regression parameters and
         mixture proportions using data simulated under the eight 
         settings of the INAR(1)-NB model (i.e. each of the scenarios from Table \ref{table:csi}).
         The simulation settings are ordered from left to right according to the class separation index with
         simulation setting $1$ having the highest class separation and simulation setting $8$
         having the lowest class separation.
         Coverage proportions are shown for simulations with $m = 200$ subjects and $m = 2,000$ 
         subjects.}}    }
\label{fig:coverage_plot}
\end{figure}

Figures \ref{fig:bias_plot} and \ref{fig:coverage_plot} 
illustrate how the underlying separation among the classes
influences other properties of estimation. Figure \ref{fig:bias_plot}
displays the absolute empirical bias for all regression coefficients and for all
eight simulation scenarios. The empirical bias for a particular parameter estimate is found
by taking the absolute value of the difference between the true parameter value and 
the mean of the simulated estimates. Figure \ref{fig:bias_plot} shows that, for high values of the CSI,
the empirical biases are packed closely to zero but spread out considerably as the CSI declines. 
Similarly, in Figure \ref{fig:coverage_plot}, we can see that the computed
confidence intervals generally give the desired $95\%$ coverage for large sample sizes ($m = 2,000$) in 
highly separated settings. However, the level of coverage and variability in coverage tends to 
deteriorate as the class separation decreases. 
\subsection{Poisson Outcomes with Normal Random Effects}
\label{ss:poisson_sims}
To evaluate the performance of our proposed fitting procedure under model
misspecification, we performed several simulations involving latent class
models where within each class we generate data under a generalized linear
mixed model with Poisson responses and Normal random effects.  The
motivation for this simulation exercise is to assess how well our method
classifies individuals to latent classes when this alternative model holds
rather than the AR(1) model.  Because we can compute the class separation
index for any Poisson-Normal model via (\ref{eq:csi}), comparing the EED
obtained from our estimated posterior probabilities to the CSI will
provide an indication of how well class assignments are recovered under
model misspecification.

\begin{figure}
\centering
     \includegraphics[width=5.5in,height=2.8in]{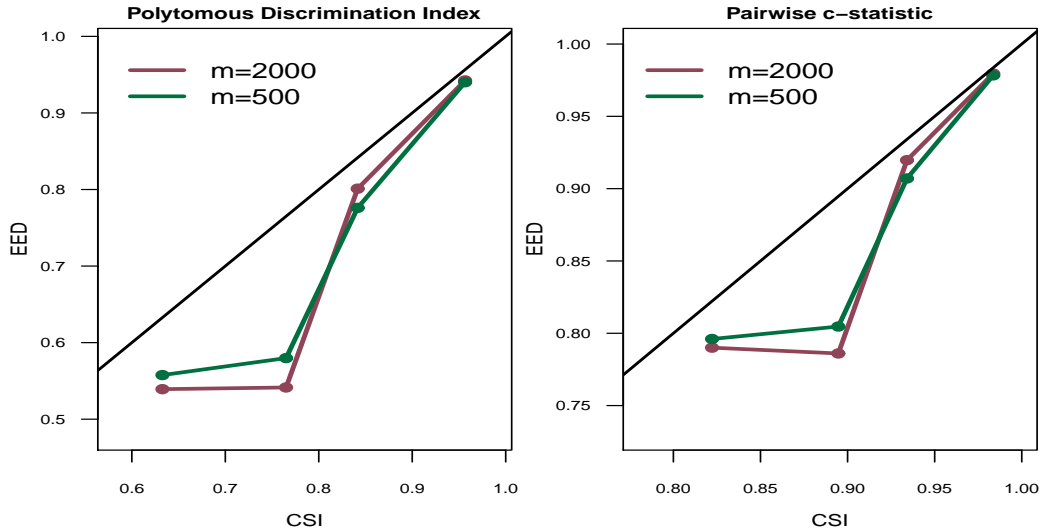}
     \caption{ {\small{Expected empirical discrimination (EED) and Class
           Separation Indices (CSI) for Poisson outcomes with Normal
           random effects models.  The values of the EED are shown when
           class assignment is based on fitting the INAR(1)-NB model with
           $m=500$ and $m=2,000$ subjects.  Fifty replications
           were used for each simulation setting and choice of $m$.}}  }
\label{fig:poisson_disc}
\end{figure}
In these simulations, conditional on a subject-specific random
slope, intercept, and class label, the response of each subject was generated
according to a Poisson distribution. In particular, for subject $i$ in 
class $c$, the $j^{th}$ response was generated as $Y_{ij} \sim \textrm{Poisson}(\lambda_{ij}^{c})$
where $\log(\lambda_{ij}^{c}) = \beta_{0}^{c} + \beta_{1}^{c}t_{ij} + a_{i0} + a_{i1}t_{ij}$, $j=1,\ldots,T$,
and where $(a_{i0},a_{i1})$ are jointly Normal random variables with
$a_{i0} \sim N(0,\sigma_{c0}^{2} + \frac{(T-1)^{2}\sigma_{c1}^{2}}{4})$, $a_{i1} \sim N(0,\sigma_{c1}^{2})$, and
$\mbox{Cov}(a_{i0},a_{i1}) = -[(T-1)\sigma_{c1}^{2}]/2$. This
means that the mean trajectories are quadratic on the log scale, viz,
\begin{equation}
\log\big( E(Y_{ij}|Z_{i} = c) \big) = \beta_{0}^{c} + \frac{\sigma_{c0}^{2}}{2}
+ \frac{ (T-1)^{2}\sigma_{c1}^{2} }{8} + \Big[ \beta_{1}^{c} - \frac{(T-1)\sigma_{c1}^{2}}{2} \Big]t_{ij} 
+ \frac{\sigma_{c1}^{2}}{2}t_{ij}^{2}.
\label{eq:poisson_quadratic}
\end{equation}
As in the simulations of Section \ref{ss:arsims}, we used four latent classes $c=1,\ldots,4$ for each simulation setting.
In total, we considered four simulation settings: one with eight time points and the
other three having five time points. In each of these, the parameters were chosen so that
the mean trajectories were similar to those in Scenario I and Scenario II. The parameter values
used for each of these four simulations settings are provided in the appendix.
For each setting of the Poisson-Normal model and each simulation replication, 
we fit a four-class integer autoregressive
model assuming that, within each class, the mean trajectory $\bmu^{c}$ was quadratic on the log scale.

The values of the class separation index and expected discrimination shown
in Figure~\ref{fig:poisson_disc} suggest that our procedure is fairly
robust to this form of misspecification. In particular, comparison of
Figure~\ref{fig:poisson_disc} to Figure~\ref{fig:discrimination} indicates
that the difference between the expected discrimination and the CSI does
not suffer greatly under model misspecification for the range of
alternative models and CSI values considered here. In addition, the high expected
discrimination in settings with a high CSI signals that our procedure will still
perform well under misspecification as long as the underlying classes are
clearly separated.

\begin{figure}
\centering
     \includegraphics[width=5in,height=2.8in]{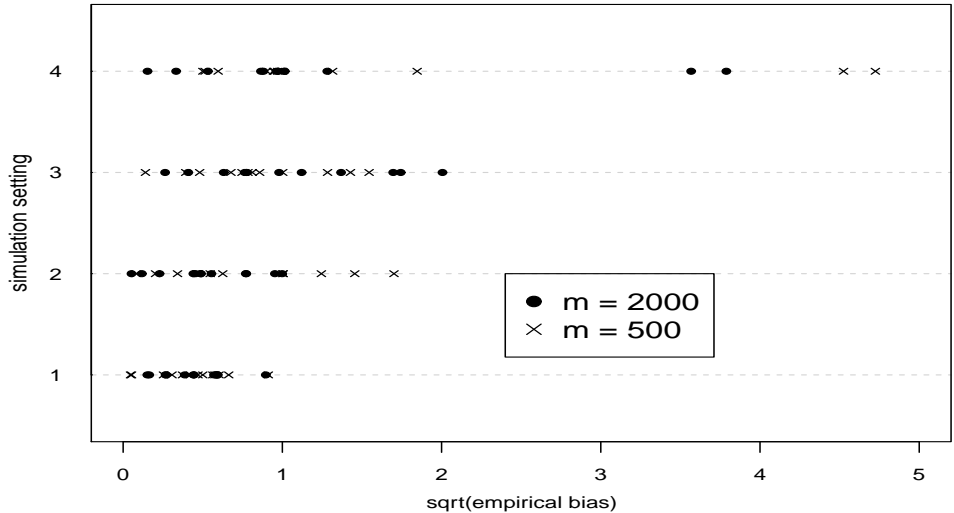}
     \caption{ {\small{Empirical bias for the regression coefficients when
           fitting an INAR(1)-NB model to data simulated under the four settings
           of the Poisson-Normal model. Computation of the bias uses the
           fact that the mean is quadratic on the log-scale (see
           (\ref{eq:poisson_quadratic})).  The simulation settings are
           ordered according to the class separation index with simulation
           setting $1$ having the highest class separation and simulation
           setting $4$ having the lowest class separation.}}
}
\label{fig:poisson_bias}
\end{figure}
We are also interested in the degree to which we recover the latent
trajectories in the Poisson-Normal model. Because of the form of the
latent trajectories in (\ref{eq:poisson_quadratic}), we can examine
the empirical bias of the regression coefficient estimates obtained from
the fit of the integer autoregressive model. Figure
(\ref{fig:poisson_bias}) displays the empirical bias of the regression
coefficient estimates across the four settings of the Poisson-Normal
simulations.  Although these results are not directly comparable to those
in Figure \ref{fig:bias_plot} due to the different number of
replications and different mean model, Figure \ref{fig:poisson_bias}
suggests that we recover the mean model reasonably well for settings with
a high class separation index.

\section{Application to CNLSY Data}
\label{s:cnlsy}
\subsection{Description and Model Fitting}
\label{ss:cnlsy_describe}
In this section, we consider data collected on Children of the National Longitudinal
Study of Youth (CNLSY). The National Longitudinal Study of Youth (NLSY79) is a longitudinal study
initiated in 1979 on a nationally representative sample of young adults. In 1986, the NLSY
launched a survey of children of female subjects from the original NLSY79 cohort.
Assessments of the children of the NLSY were performed biennially, and
in each assessment, mothers were asked to respond to a series of questions regarding each of their
children's behavior and environment.

Although the mothers were asked to respond on a wide variety of measures, our focus
lies in the severity of behavioral problems as measured by the ``Behavior Problems Index'' (BPI)
in early childhood and by the number of delinquent acts committed during the adolescent years.
The BPI is recorded for children ages $4-13$ and is constructed by asking the mother to rate
her child on a scale of 0 to 2 on each item from a list of seven common behavioral problems.
Consequently, this yields a BPI value which is an integer between $0$ and $14$.
Starting at the age of $10$, the children were also asked to report the number
of delinquent acts they committed during the past year.
From age 14 onward, the mothers were no longer asked to respond to the BPI questions,
leaving only the self-reported delinquency counts as a measure of behavioral conduct for children
older than 13.

%
\begin{table}[h]
\caption{   {\small{Summary statistics from the CNLSY data. In total, these data contain $9,626$ subjects
each of which was surveyed biennially over the ages 4 to 16 (or 5 to 17). 
For the age groups 4-5, 6-7, and 8-9,
the counts are solely from the behavioral problems index (BPI). For age groups 10-11 and 12-13, the
counts represent the sum of the mother-reported BPI and the child-reported number of delinquent acts. For
age groups 14-15 and 16-17, the counts are solely from the self-reported number of delinquent acts.}}   }
\begin{center}
{\renewcommand{\arraystretch}{1.0} \begin{tabular}{lllllllr}
\toprule
Child Ages    & Mean & 10\% & 25\% & 50\% & 75\% & 90\% & Max \\
\midrule
4-5       &  2.35 &  0 &  1  &  2 &  4 & 6   & 14   \\
6-7       &  2.12 &  0 &  0  &  2 &  3 & 5   & 14   \\
8-9       &  2.15 &  0 & 0   &  2 &  3 & 5   & 14   \\
10-11     &  3.46 &  0 & 1   &  3 &  5 & 8   & 19   \\
12-13     &  3.74 &  0 & 1   &  3 &  5 & 8   & 20  \\
14-15     &  1.36 &  0 & 0   &  1 &  2 & 4   & 7   \\
16-17     &  1.37 &  0 & 0   &  1 &  2 & 4   & 7 \\
\bottomrule
\end{tabular}
}
\end{center}
\label{table:cnlsy_summary}
\end{table}

To model the evolution of behavioral problems throughout childhood and adolescence,
we added the BPI and the delinquency counts into a single response variable at each time point.
For children less than 10, we used the BPI as the only response,
and for children aged 10-13 we took the response to be the sum of the delinquency counts
and the BPI. For children older than 13 the response is simply the delinquency
counts. To account for this methodological feature of the measurement process, we added 
a dummy variable for the time points corresponding to
the age groups 10-11 and 12-13. In addition, because the mean of the delinquency counts
is lower than the mean of the BPI (see Table \ref{table:cnlsy_summary}), we added another dummy variable for children
older than $13$.

We modeled the subject-specific trajectories $\bmu_{i}^{c} = (\mu_{i1}^{c},\ldots,\mu_{in_{i}}^{c})$, with
$\mu_{ij}^{c}$ denoting the mean response of subject $i$ at time $t_{ij}$
conditional on belonging to class $c$, as
\begin{equation}
\log(\mu_{ij}^{c}) = \beta_{0}^{c} + \sum_{k=1}^{3}{\beta_{k}^{c}B_{k}(t_{ij})} + \beta_{4}^{c}\big( \mathbf{1}\{t_{ij}=4\} + \mathbf{1}\{t_{ij}=5\}\big)
+ \beta_{5}^{c}\mathbf{1}\{t_{ij} \geq 6\}.
\label{eq:CNLSYregression}
\end{equation}
In (\ref{eq:CNLSYregression}), $\{ B_{k}(\cdot)\}_{k=1}^{3}$ are the
B-spline basis functions of degree $3$ with no interior knots. We coded
the time variable as follows: $t_{ij} = 1$ for children ages 4-5,
$t_{ij} = 2$ for children ages 6-7 with the remaining five time points
coded similarly. To handle subjects that had missing responses, we assumed
that the correlation only depended on the order of the observed
responses. For example, if subject $i$ was observed at times $1$, $2$, and
$4$ with a missing value at time point $3$, then we would have
$\corr_{\theta_{c}}(y_{i4},y_{i2}) = \alpha_{c}$ and
$\corr_{\theta_{c}}(y_{i4},y_{i1}) = \alpha_{c}^{2}$.

\begin{figure}
\centering
     \includegraphics[width=6in,height=3.5in]{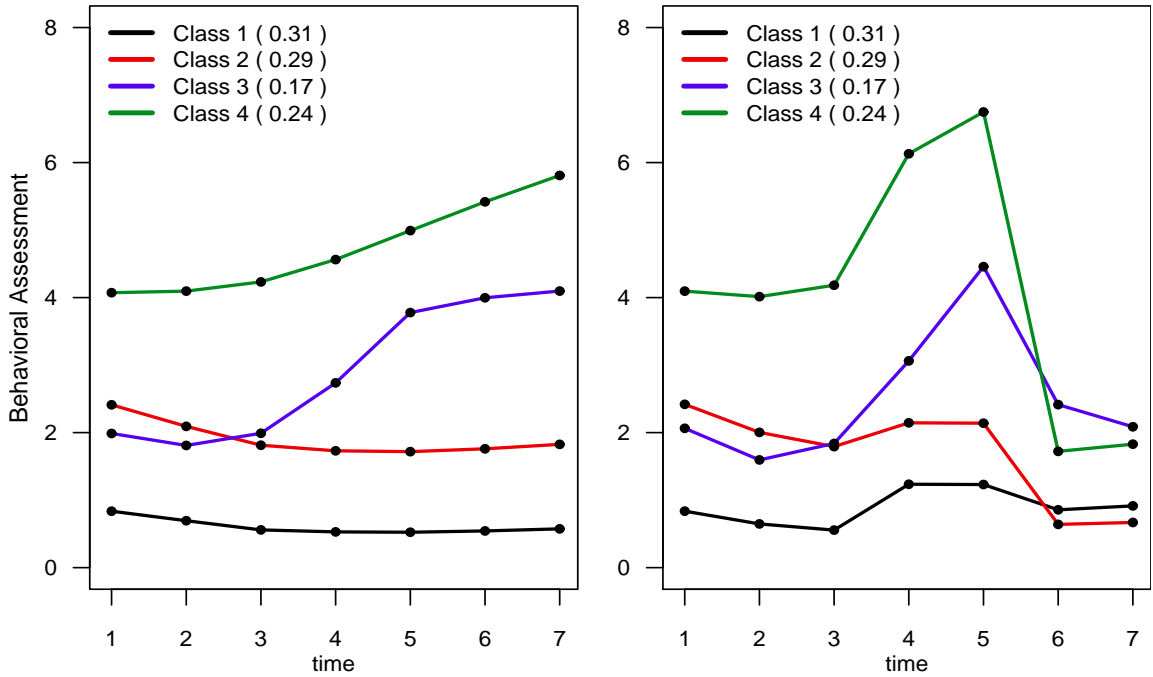}
\caption{   {\small{Estimated trajectories for the CNLSY data assuming four latent classes. The terms
 in parentheses represent estimated class-membership proportions. The left panel displays
 the estimated trajectories adjusted for the different measurement scales
 used at different time points; these can be thought of as the fitted latent trajectories
 on the mother-reported BPI measurement scale. More specifically, the fitted curves in the left panel
 do not include the time indicators present in equation (\ref{eq:CNLSYregression}) while the right-hand panel displays
 the fitted trajectories associated with the full model in equation (\ref{eq:CNLSYregression}).}}   }
\label{fig:fittedcurves}
\end{figure}
The CNLSY provides sampling weights which are estimates of how many
individuals in the population are represented by a particular subject in
the data. Hence, the sampling weights
reflect the inverse probability that a subject was included in the sample.
To account for this aspect of the sampling process, we fitted
latent class models where, within each iteration of the algorithm, we
solve a weighted version of (\ref{eq:weightedgee}) and evaluate
convergence with a weighted version of the estimating function
$G(\bTheta,\bpi)$ defined in Section \ref{ss:stderrs}.  This modified
version of the fitting procedure that accounts for sampling weights is
described in more detail in the appendix.

We applied our procedure to the CNLSY data varying the number of classes
from $3$ to $6$ where, in each case, we modeled the mean trajectories with
(\ref{eq:CNLSYregression}). As is advisable in latent class modeling applications, 
for each choice of the number of classes, we repeated the estimation procedure 
with $20$ different starting values; in each case, the weighted log-likelihood converged to the same maximum
value for at least two runs.  When comparing the best runs of these four
different models (i.e., the 3 to 6 class models), the four-class model
possessed the lowest weighted BIC though the four and five-class models had nearly
identical values of the weighted BIC. In particular, the best values 
of the weighted BIC for the 3 to 6 class models were $161,751.3$, 
$161,651.2$, $161,651.4$, and $161,681.0$ respectively. The fitted trajectories and
estimated class-membership proportions from the four-class solution are
displayed in Figure \ref{fig:fittedcurves}, and here we labeled the four
classes to roughly reflect increasing levels of conduct disorder severity
with Class 1 (4) denoting the least (most) severe class.  The right-hand panel of Figure \ref{fig:fittedcurves}
displays the estimated mean curves $(\hat{\mu}_{i1},\ldots,\hat{\mu}_{i7}^{c})$ including all the predictors
from (\ref{eq:CNLSYregression}). The left-hand panel displays the mean curves obtained 
by excluding the time point indicators $\mathbf{1}\{t_{ij}=4 \mbox{ or }t_{ij}=5\}$ and
$\mathbf{1}\{t_{ij} \geq 6\}$, and is intended to reflect population
trajectories in terms of the BPI only.

\begin{table}[h]
\caption{   {\small{ Weighted cross-tabulation of random class assignment and maternal age at
the birth of the child using only male subjects, and weighted cross-tabulation of class assignment 
and of ever being convicted of a crime during ages $14-18$ using only male subjects.
The class assignments were obtained by using the estimated posterior probabilities
to randomly assign each subject to one of the four latent classes. 
The random assignment procedure was repeated $1,000$ times, and the results were averaged.
In the top table, we display in parentheses class proportions conditional on maternal age 
while in the bottom table we show conviction proportions conditional on class.
}}   }
\begin{center}
{\renewcommand{\arraystretch}{1.0} 
\begin{tabular}{lllll}
\toprule
       & Class 1 & Class 2 &  Class 3 &  Class 4  \\
\midrule
maternal age $< 20$    &   $34.2$ ($0.178$) &  $44.8$ ($0.233$)  &  $40.2$ ($0.209$)   &  $73.1$ ($0.380$)   \\
maternal age $\geq 20$ &  $968.7$ ($0.284$) & $941.7$ ($0.276$)  &  $581.7$ ($0.171$)  &  $917.6$ ($0.269$)    \\
\bottomrule
\end{tabular}
}
\end{center}

\begin{center}
\begin{tabular}{lllll}
\toprule
       & Class 1 & Class 2 &  Class 3 &  Class 4  \\
\midrule
ever convicted-yes      &  $18.7$ ($0.031$)  &  $16.5$ ($0.025$)   &  $42.6$ ($0.105$)  &  $102.9$ ($0.144$)   \\
ever convicted-no       &  $588.0$ ($0.969$) &  $646.8$ ($0.975$) &  $363.1$ ($0.895$)  &  $609.4$ ($0.856$)   \\
\bottomrule
\end{tabular}
\end{center}
\label{table:ageconvict_tables}
\end{table}

Additional variables in the CNLSY such as gender and criminal history allow us to
examine associations between these variables and the latent classes identified in the four-class solution. 
Because this is an analysis that is relevant to the domain area of developmental 
psychopathology, these associations are critical for substantive validation of 
the four extracted classes. To investigate these associations, we randomly assigned
each subject to one of the four identified classes using the estimated posterior
probabilities of class membership and cross-tabulated these assignments with
other variables in the CNLSY. Table \ref{table:ageconvict_tables} displays a
weighted contingency table of the random class assignment and  
maternal age at birth ($< 20$ years old, or $\geq 20$ years old) only for male subjects,
and the resulting frequencies seem to support the validity of the four identified classes as
the proportion of subjects in classes 1 or 2 with maternal age $\geq 20$ is considerably
higher than the proportion of subjects in classes 1 or 2 with maternal age $< 20$.
Table \ref{table:ageconvict_tables} also shows a weighted cross tabulation
of class assignment and of ever being convicted of a crime between ages $14-18$, and  
as demonstrated by this contingency table, the prevalence of criminal outcomes in classes 3 and 
4 is substantially higher than in classes 1 and 2. Moreover, the frequency of criminal outcomes
in those assigned to class 4 is considerably greater than that of the other three classes.

\subsection{Diagnostics}
\label{ss:diagnostics}
Whereas there are a variety of aspects of the model which we could assess, our interest here lies in checking whether
our specification of the within class distribution is reasonable. In particular, we 
are especially interested in examining if the assumed within class correlation structure seems to hold in the
CNLSY data; that is, we want to check if $\corr(y_{ik},y_{ij}|Z_{i}=c)$ decays exponentially over time. 
If the class labels were known, we could check this by directly stratifying subjects
into their respective classes and computing the desired correlations.
We mimic this process by using each subject's estimated posterior probabilities to randomly assign
each subject to one of the latent classes. Such a diagnostic approach, where within-class quantities
are checked by sampling from the estimated posterior probabilities, is similar to the
procedure discussed in \cite{BandeenRoche:1997}. As shown in that paper,
this procedure is valid for detecting departures from the model if the assumed latent class model is
indeed false.
\begin{figure}
\centering
     \includegraphics[width=5in,height=3in]{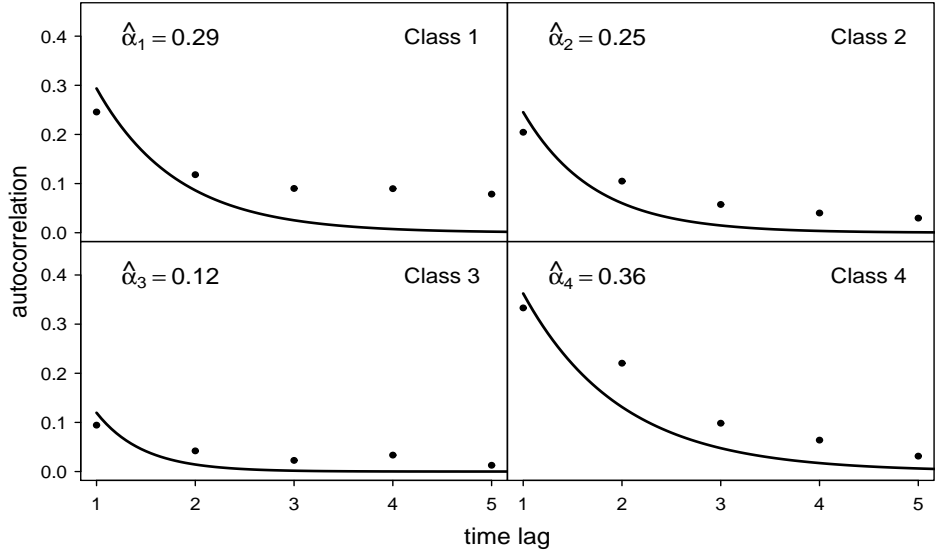}
\caption{   {\small{Sample autocorrelation functions (weighted) obtained by using the estimated posterior probabilities
 of class membership to randomly assign each subject to one 
 of the four latent classes. The random assignment procedure was repeated $1,000$ times;
 the displayed autocorrelation values represent the average autocorrelation values
 from these replications.}}   }
\label{fig:diagnosis}
\end{figure}

\begin{figure}
\centering
     \includegraphics[width=5in,height=3in]{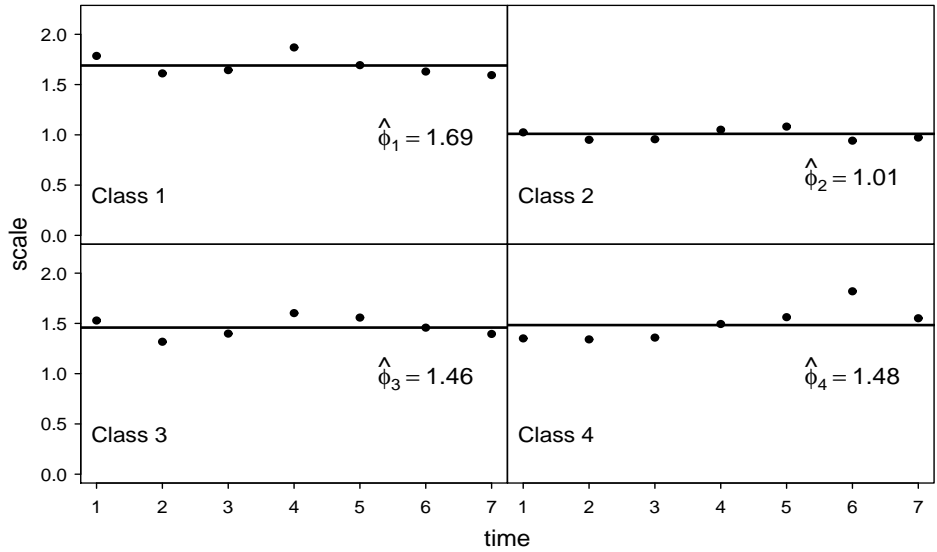}
\caption{   {\small{ Sample overdispersion (weighted) obtained by using the estimated posterior probabilities
 of class membership to randomly assign each subject to one 
 of the four latent classes. This random assignment procedure was repeated $1,000$ times;
 the displayed overdispersion values represent the average overdispersion values
 from these replications.
 }}   }
\label{fig:scale_diagnosis}
\end{figure}

Estimated autocorrelation functions obtained from the random stratification procedure
described above are displayed in Figure \ref{fig:diagnosis} with the same class labels
as in Figure \ref{fig:fittedcurves}. For each random stratification, we used the subject-specific sampling weights
to compute weighted estimates of the autocorrelation 
and then averaged the results over $1,000$ replications of the random stratification process.
The autocorrelation plots in Figure \ref{fig:fittedcurves}
show that the AR(1) structure seems somewhat plausible for the CNLSY data in that the
within-class correlations decay substantially over time. 
However, for most classes, the correlations do not seem to decay quite as quickly
as would occur under an AR(1) assumption.  
A similar diagnostic plot for the scale parameters $\phi_{c}$ is shown in Figure \ref{fig:scale_diagnosis}.
This shows estimates of overdispersion for each time point obtained through repeatedly performing random class assignment.
The plots in Figure \ref{fig:scale_diagnosis} suggest that the original estimates of $\phi_{c}$ are reasonable and 
that the level of overdispersion is relatively constant over time.

\section{Discussion}
\label{s:discussion}
We have presented a method for performing latent class analysis on longitudinal count data.
The motivation behind this work has been to express many of the core features of latent class models
or growth mixture models in a straightforward, computationally tractable
framework that will improve computational performance and model
identifiability, and that will perform well even if the true data generating
mechanism is the popular growth mixture model.
The autoregressive structure used in our model serves both as a natural way
to model dependence over time and to achieve clearer separation of
correlation due to the repeated-measurements structure and correlation due
to the latent class structure.  
In terms of computation, one of the chief advantages of this approach is the availability of the subject-specific
likelihood in closed form; this simplifies computation considerably at least when compared with procedures that 
employ random effects and which require the use of numerical integration procedures within each class. 
In addition, because computational efficiency has been a primary motivation,
we described a quasi-EM approach which we found to be especially useful in this setting.

We also have offered novel notions of class separation inherent in the
data and of a given modeling procedure to recover that level of
discrimination. Based on model classification indices such as those used
in classification regression models, the oracle-based CSI quantifies the
degree to which the information in the data can correctly classify
subjects given the correct model and parameters.  The EED, by contrast,
quantifies the degree to which a given procedure and sample size can
correctly classify subjects; the EED can be compared to the CSI, the
latter acting as an upper bound. We found excellent classification
performance by applying our modeling procedure to Poisson-normal random
effects latent class model data. Our model performed very nearly as well
as it did when our model was the data generating mechanism with
comparable CSI values, although the range of alternative models considered
here is quite limited and investigating other forms of misspecification
would certainly be worthwhile.

\appendix
\section{Appendix}

\subsection{Transition Function}
The transition function $p(y_{ij}|y_{i,j-1},\mathbf{X}_{i},\btheta_{c}) = p_{i}(y_{ij}|y_{i,j-1};\btheta_{c})$
can be expressed as the convolution of a beta-binomial distribution and a negative binomial distribution
\begin{equation}
p_{i}(y_{ij}|y_{i,j-1};\btheta_{c}) = \sum_{k=0}^{\min(y_{ij},y_{i,j-1})}{ f_{i}(y_{ij} - k)g_{i}(k)}.
\label{eq:convolution}
\end{equation}
Above, $f_{i}(\cdot)$ represents the probability mass function for a beta-binomial distribution
with parameters $\Big( y_{ij},\alpha_{c}\mu_{i,j-1}^{c}/\gamma_{c}, (1 - \alpha_{c})\mu_{i,j-1}^{c}/\gamma_{c} \Big)$
\begin{equation}
f_{i}(k) = \frac{\Gamma(y_{i,j-1} + 1)\Gamma(\eta_{i,j-1}^{c})\Gamma(k + \alpha_{c})\eta_{i,j-1}^{c})\Gamma(y_{i,j-1} + (1 - \alpha_{c})\eta_{i,j-1}^{c})}
{\Gamma(k+1)\Gamma(y_{i,j-1} - k + 1)\Gamma(\alpha_{c}\eta_{i,j-1}^{c})\Gamma((1 - \alpha_{c})\eta_{i,j-1}^{c})\Gamma(\eta_{i,j-1}^{c} + y_{i,j-1})}
\end{equation}
and $g_{i}(\cdot)$ represents the probability mass function for a negative binomial distribution with mean
$\mu_{ij}^{c}(1 - \alpha_{c})$ and variance $\mu_{ij}^{c}(1 - \alpha_{c})(1 + \gamma_{c})$
\begin{equation}
g_{i}(k) = \frac{\Gamma\{ k + (1 - \alpha_{c})\mu_{ij}^{c}/\gamma_{c} \}}{\Gamma\{k+1\}\Gamma\{k + (1 - \alpha_{c})\mu_{ij}^{c}/\gamma_{c}\} }
\Big( \frac{\gamma_{c}}{1 + \gamma_{c}}  \Big)^{(1 - \alpha_{c})\mu_{ij}^{c}/\gamma_{c}}
\Big( \frac{\gamma_{c}}{1 + \gamma_{c}} \Big)^{k}.
\end{equation}
By applying (\ref{eq:convolution}), the transition function can then be written as
\begin{eqnarray}
p_{i}(y_{ij}|y_{i,j-1};\btheta_{c}) &=& \frac{\gamma_{c}^{y_{i,j-1}}\Gamma(y_{i,j-1} + 1)\Gamma(\eta_{i,j-1}^{c}) }
{ \Gamma(\lambda_{ij}^{c})\Gamma(\eta_{i,j-1}^{c} - \lambda_{ij}^{c})\Gamma(\eta_{ij}^{c} - \lambda_{ij}^{c})
\Gamma(\eta_{i,j-1}^{c} + y_{i,j-1})(1 + \gamma_{c})^{y_{i,j-1} + \eta_{ij}^{c} - \lambda_{ij}^{c}}  }   \nonumber \\
& & \times \sum_{k=0}^{\min(y_{ij},y_{i,j-1})}{  \Big( \frac{1 + \gamma_{c}}{\gamma_{c}} \Big)^{k}
\frac{\Gamma(\lambda_{ij}^{c})\Gamma(\eta_{i,j-1}^{c} - \lambda_{ij}^{c} + y_{i,j-1} - k)\Gamma(\eta_{ij}^{c} - \lambda_{ij}^{c} + y_{ij} - k)}
{\Gamma(y_{i,j-1} - k + 1)\Gamma(y_{ij} - k + 1)\Gamma(k+1)}  }, \nonumber
\end{eqnarray}
where $\lambda_{ij}^{c} = \alpha_{c}\sqrt{\mu_{ij}^{c}\mu_{i,j-1}^{c}}/\gamma_{c}$, $\eta_{ij}^{c} = \mu_{ij}^{c}/\gamma_{c}$,
and $\eta_{i,j-1}^{c} = \mu_{i,j-1}^{c}/\gamma_{c}$.

\subsection{Parameter Initialization}
The parameter initialization procedure is detailed below
\begin{itemize}
 \item[\textbf{1.}] 
 Choose $K$ ``cluster centers'' for the regression coefficients $(\bbeta_{1},\ldots,\bbeta_{K})$.
 This is done by randomly selecting $K$ subject and fitting a separate Poisson regression for
 each subject.
 \item[\textbf{2.}] 
 Assign each subject to one of the $K$ classes through $S_{i} = \underset{c}{\operatorname{argmin }}\{D(\mathbf{y}_{i};\bbeta_{c}) \}$.
 Here, $D(\mathbf{y}_{i};\bbeta_{c}) = -2\log L(\bbeta_{c}|\mathbf{y}_{i})$ is the usual deviance
 associated with a Poisson regression. Namely,
 $D(\mathbf{y}_{i};\bbeta_{c}) = 2\sum_{j=1}^{n_{i}}{\Big( \mu_{ij}^{c} - y_{ij}\log(\mu_{ij}^{c}) + \log(y_{ij}!) \Big)}$.
 \item[\textbf{3.}] 
 Using this hard assignment of subjects to clusters, compute a new value of $\bbeta_{c}$ by fitting
 a Poisson regression for the subjects in the set $\mathcal{S}_{c} = \{i: S_{i}=c \}$. Do this
 for each cluster $c=1,\ldots,K$.
 \item[\textbf{4.}] 
 Repeat steps (2)-(3) twice.
 \item[\textbf{5.}] 
 Compute the mixture proportions through $\pi_{c} = \tfrac{1}{m}\sum_{i=1}^{m}{\mathbf{1}\{ S_{i} = c \} }$,
 and compute each $\btheta_{c} = (\bbeta_{c},\alpha_{c},\gamma_{c})$ by solving $\sum_{i \in \mathcal{S}_{c}}{U_{i}(\btheta_{c})} = \mathbf{0}$,
 where $U_{i}(\cdot)$ is as described in equation (\ref{eq:maingee}).
\end{itemize}

\subsection{Estimation with Sampling Weights}
Suppose $v_{i}$, $i = 1,\ldots,m$ are sampling weights such that
$v_{i}$ is proportional to the inverse probability that subject $i$
is included in the sample. We compute initial estimates $(\bTheta^{(0)},\bpi^{(0)})$
in the same way as the unweighted case. Given estimates $(\bTheta^{(k)},\bpi^{(k)})$,
we produce updated estimates $(\bTheta^{(k+1)}, \bpi^{(k+1)} )$ 
in the $(k+1)^{st}$ step through the following process.
\begin{itemize}
\item
Update $\btheta_{c} = (\bbeta_{c},\alpha_{c},\gamma_{c})$ by solving
\begin{equation}
\sum_{i=1}^{m}{ v_{i} W_{ic}( \bTheta^{(k)}, \bpi^{(k)}) U_{i}(\btheta_{c}) } = \mathbf{0},
\end{equation}
where $U_{i}(\cdot)$ is the estimating function described in equation (\ref{eq:maingee}).
\item
Update $\pi_{c}$ through
\begin{equation}
\pi_{c}^{(k+1)} = \frac{ \sum_{i=1}^{m}{ v_{i}W_{ic}(\bTheta^{(k)},\bpi^{(k)}) } }{\sum_{i=1}^{m}{ v_{i} } }
\end{equation}
\end{itemize}
We determine convergence by stopping when the weighted estimating function
$\sum_{i=1}^{m}{ v_{i}G_{i}(\bTheta,\bpi) }$ is sufficiently close to zero.

\subsection{Simulation Parameter Values}
In Scenarios I and II of the INAR simulations, we define the mean curves
through $\log(\mu_{ij}^{c}) = \beta_{0}^{c} + \beta_{1}^{c}t_{ij}$. 
The values of $(\beta_{0}^{c},\beta_{1}^{c})$ used for Scenarios I and II 
are given in Table \ref{table:coefsimvalues}.
\begin{table}[h]
\caption{Values of the regression coefficients in Scenarios I and II.}
\begin{center}
{\renewcommand{\arraystretch}{1.8} 
    \begin{tabular}{l| llll | llll   }
     \toprule
      \multicolumn{1}{c}{}& \multicolumn{4}{ c| }{Scenario I} & \multicolumn{4}{ |c }{Scenario II} \\
     \cline{1-9}
     \multicolumn{1}{c}{}& c=1 & c=2 & c=3 & c=4 & c=1 & c=2 & c=3 & c=4 \\
    \hline
    $\beta_{0}^{c}$ & -0.4 & 1.5 & 0.0 & 1.4 & -0.4 & 1.4 & 0.0 & 1.2  \\
    $\beta_{1}^{c}$ & -0.1 & -0.7 & 0.65 & 0.0 & -0.1 & -1.0 & 0.9 & 0.0 \\
    \bottomrule
    \end{tabular}
}
\end{center}
\label{table:coefsimvalues}
\end{table}

The Poisson-Normal simulations use four simulation settings. One of these settings 
(the setting with the largest class separation index) has eight time points with
time points $t_{ij} = j/9$ for $j=1,\ldots,8$. The other three settings have five time points with
$t_{ij} = j/6$ for $j=1,\ldots,5$. Each setting utilizes the 
parameters $(\beta_{0}^{c},\beta_{1}^{c},\sigma_{c0}^{2},\sigma_{c1}^{2},\pi_{c})$ for $c = 1,\ldots,4$.
The values of these parameters for each simulation setting (as indexed by the value of the CSI)
are shown in table \ref{table:poissonpars} 
\begin{table}[h]
\caption{Parameter values used for the Poisson-Normal simulations. The Class Separation Index (CSI)
        shown for each simulation setting was computed with the polytomous discrimination index.}
\begin{center}
{\renewcommand{\arraystretch}{1.4} 
    \begin{tabular}{l| llll    }
     \toprule
     \multicolumn{1}{c}{}& CSI=0.957 & CSI=0.842 & CSI=0.765 & CSI = 0.633  \\
    \hline
    $\beta_{0}^{1}$ & -0.90 & -0.90 & -0.90 & -9.00   \\
    $\beta_{1}^{1}$ & -0.35 & -0.35 & -0.35 & -0.35  \\
    $\sigma_{10}^{2}$ & 0.30 & 0.40 & 0.85 & 1.50   \\
    $\sigma_{11}^{2}$ & 0.125 & 0.20 & 0.50 & 0.70  \\
    $\pi_{1}$ & 0.50 & 0.50 & 0.50 & 0.50  \\
    \hline
    $\beta_{0}^{2}$ & 1.55 & 1.55 & 1.55 & 1.3   \\
    $\beta_{1}^{2}$ & -2.10 & -2.10 & -2.10 & -2.10  \\
    $\sigma_{20}^{2}$ & 0.08 & 0.15 & 0.35 & 1.00   \\
    $\sigma_{21}^{2}$ & 0.05 & 0.075 & 0.25 & 0.50  \\
    $\pi_{2}$ & 0.25 & 0.25 & 0.25 & 0.25  \\
    \hline
    $\beta_{0}^{3}$ & -0.40 & -0.65 & -0.65 & -0.7   \\
    $\beta_{1}^{3}$ & 1.90 & 2.00 & 2.00 & 1.75  \\
    $\sigma_{30}^{2}$ & 0.10 & 0.20 & 0.60 & 1.25   \\
    $\sigma_{31}^{2}$ & 0.06 & 0.075 & 0.25 & 0.55  \\
    $\pi_{3}$ & 0.15 & 0.15 & 0.15 & 0.15  \\
    \hline
    $\beta_{0}^{4}$ & 1.40 & 1.25 & 1.25 & 1.00   \\
    $\beta_{1}^{4}$ & -0.05 & -0.05 & -0.05 & -0.05  \\
    $\sigma_{40}^{2}$ & 0.06 & 0.10 & 0.35 & 1.00   \\
    $\sigma_{41}^{2}$ & 0.04 & 0.04 & 0.30 & 0.45  \\
    $\pi_{4}$ & 0.10 & 0.10 & 0.10 & 0.10  \\
    \bottomrule
    \end{tabular}
}
\end{center}
\label{table:poissonpars}
\end{table}

\bibliographystyle{biom.bst}
\nocite{*}
\bibliography{inar_arxiv150123}

\end{document}